\documentclass[twocolumn,useAMS,usenatbib]{mn2e}
\usepackage{graphicx}
\usepackage{natbib}
\usepackage{bm}
\usepackage{amsfonts}
\usepackage{latexsym}
\usepackage[latin1]{inputenc}
\usepackage{amsmath}
\usepackage{epsfig}
\usepackage{amsbsy}
\usepackage{psfrag}
\usepackage{mnfixes}

\newcommand{\mtb}[1]{\mathbf{#1}}

\newcommand{\mtv}[1]{\textbf{\emph{#1}}}

\newcommand{\p}{\mathrm p}

\newcommand{\veps}{\varepsilon}

\def \nn  {\nonumber}

\def \veps{\varepsilon}
\def\jnl@style{\rm}
\def\aaref@jnl#1{{\jnl@style#1}}
\def\aaref@jnl#1{{\jnl@style#1}}

\def\aj{\aaref@jnl{AJ}}                   % Astronomical Journal
\def\apj{\aaref@jnl{ApJ}}                 % Astrophysical Journal
\def\apjl{\aaref@jnl{ApJ}}                % Astrophysical Journal, Letters
\def\apjs{\aaref@jnl{ApJS}}               % Astrophysical Journal, Supplement
\def\apss{\aaref@jnl{Ap\&SS}}             % Astrophysics and Space Science
\def\aap{\aaref@jnl{A\&A}}                % Astronomy and Astrophysics
\def\aapr{\aaref@jnl{A\&A~Rev.}}          % Astronomy and Astrophysics Reviews
\def\aaps{\aaref@jnl{A\&AS}}              % Astronomy and Astrophysics, Supplement
\def\mnras{\aaref@jnl{MNRAS}}             % Monthly Notices of the RAS
\def\prd{\aaref@jnl{Phys.~Rev.~D}}        % Physical Review D
\def\prl{\aaref@jnl{Phys.~Rev.~Lett.}}    % Physical Review Letters
\def\qjras{\aaref@jnl{QJRAS}}             % Quarterly Journal of the RAS
\def\skytel{\aaref@jnl{S\&T}}             % Sky and Telescope
\def\ssr{\aaref@jnl{Space~Sci.~Rev.}}     % Space Science Reviews
\def\zap{\aaref@jnl{ZAp}}                 % Zeitschrift fuer Astrophysik
\def\nat{\aaref@jnl{Nature}}              % Nature
\def\aplett{\aaref@jnl{Astrophys.~Lett.}} % Astrophysics Letters
\def\apspr{\aaref@jnl{Astrophys.~Space~Phys.~Res.}} % Astrophysics Space Physics Research
\def\physrep{\aaref@jnl{Phys.~Rep.}}      % Physics Reports
\def\physscr{\aaref@jnl{Phys.~Scr}}       % Physica Scripta

\title[low-T/W instability]{Merger-inspired rotation laws and the low-T/W instability in neutron stars}

\author[A. Passamonti and N. Andersson]
{A. Passamonti$^1$\thanks{E-mail:apfisica@yahoo.it} , N. Andersson$^2$ \\ \\
$^1$Via Greve 10, 00146, Roma, Italy \\
$^2$School of Mathematics and STAG Research Centre, University of Southampton, Southampton SO17 1BJ, UK}

\begin{document}

%%%%%%%%%%%%%%%%%%%%%%%%%%%%%%%%%%%%  DATE  %%%%%%%%%%%%%%%%%%%%%%%%%%%%%%%%%%%%
\date{\today}

%%%%%%%%%%%%%%%%%%%%%%%%%%%%%%% PAGE RANGE  %%%%%%%%%%%%%%%%%%%%%%%%%%%%%%%%%%%%
\pagerange{\pageref{firstpage}--\pageref{lastpage}} \pubyear{}
%%%%%%%%%%%%%%%%%%%%%%%%%%%%%  MAKETITLE  %%%%%%%%%%%%%%%%%%%%%%%%%%%%%%%%%%%%%%
\maketitle

%%%%%%%%%%%%%%%%%%%%%%%%%%%%%%% FIRST PAGE  %%%%%%%%%%%%%%%%%%%%%%%%%%%%%%%%%%%%
\label{firstpage}

%%%%%%%%%%%%%%%%%%%%%%%%%%%%%  ABSTRACT  %%%%%%%%%%%%%%%%%%%%%%%%%%%%%%%%%%%%%%%

\begin{abstract}
Implementing a family of differential rotation laws inspired by binary neutron-star merger remnants, we consider the impact of the rotation profile on the low-T/W instability. 
We use time evolutions of the linearised dynamical equations, in Newtonian gravity, to study non-axisymmetric oscillations and identify the unstable modes. 
The presence and evolution of the low-T/W instability is monitored with the canonical energy and angular momentum, while the growth time is extracted from the evolved kinetic energy. 
The results for the new rotation laws highlight similarities with the commonly considered j-constant law. The instability sets in when an oscillation mode co-rotates with the star (i.e. whenever there is a point where the mode's pattern speed matches the bulk angular velocity) and grows faster deep inside the co-rotation region.
However, the new profiles add features, like an additional co-rotation point to the problem, which affect the onset of instability. 
The rotation laws influence  more drastically the oscillation frequencies of the $l=m=2$ f-mode in fast rotating models, but affect the instability growth time 
at any rotation rate. We also identify models where the low-T/W instability appears to be triggered by inertial modes. 
We discuss to what extent the inferred qualitative behaviour is likely to be of observational relevance.  
\end{abstract}

%%%%%%%%%%%%%%%%%%%%%%%%%%%%%  Keywords  %%%%%%%%%%%%%%%%%%%%%%%%%%%%%%%%%%%%%%%%%%%
\begin{keywords}
methods: numerical -- stars: neutron -- stars: oscillation -- star:rotation.
\end{keywords}

%%%%%%%%%%%%%%%%%%%%%%%%%  SEC. I: INTRODUCTION  %%%%%%%%%%%%%%%%%%%%%%%%%%%%%%%
\section{Introduction}
%%%%%%%%%%%%%%%%%%%%%%%%%%%%%%%%%%%%%%%%%%%%%%%%%%%%%%%%%%%%%%%%%%%%%%%%%%%%%%%%

With the "simultaneous" detection of gravitational and electromagnetic signals from binary neutron stars \citep{PhysRevLett.119.161101} we truly entered  the era multimessenger astronomy. 
The gravitational-wave signal from such mergers depends on a number of physical processes operating during the 
coalescence phase, the merger and the post-merger dynamics \citep{2017RPPh...80i6901B,bernuzzi}. The main parameters that determine the fate of the post-merger remnant are the total mass and the angular momentum (although different aspects of the physics, like the matter equation of state, magnetic fields may impact significantly on the dynamics). Broadly speaking, the remnant  may continue to live as a stable compact star or undergo  collapse, potentially delayed as the system loses angular momentum, leading to the distinction between hypermassive and supramassive remnants \citep*{2000ApJ...528L..29B}.

Hypermassive neutron stars may survive longer and avoid prompt core collapse because of differential rotation. Still, 
as their mass lies above the critical mass that can be supported by uniformly rotating stars, they become unstable as dissipative processes 
smooth out the differential rotation. 
Nonlinear  simulations of hypermassive remnants from binary neutron star mergers
 have shown  that the rotation profile may be non-trivial, especially during the early postmerger phase \citep{2015PhRvD..91f4027K, 2016PhRvD..94d4060K, 2017PhRvD..95f3016C,  2017PhRvD..96d3004H, 2017PhRvD..96d3019K, 2018PhRvD..97l4039K, 2018PhRvD..98d3015E, PhysRevD.100.023005, 2019arXiv191004036D}. The results suggest that the core of the remnant  generally rotates slower than the outer layers, and these layers typically approach the Kepler velocity at larger distances, representing a disk of orbiting material \citep{2017PhRvD..96d3019K}. This profile is rather different from that commonly assumed in work on the dynamics of differentially rotating neutron stars, which tends to focus on the so-called j-constant law (the relativistic generalisation of a system with constant specific angular momentum, see for instance \citet{1986ApJS...61..479H} and \citet{Komatsu1989MNRASA, Komatsu1989MNRASB}). This then naturally leads to the question of how the differential rotation law impacts on the dynamics of the object. 

A particularly interesting aspect of this question is associated with the fact that the gravitational-wave signal may be amplified by non-axisymmetric instabilities, developing 
on a dynamical timescale. It is well known that, in relativistic stars with realistic tabulated equation of state (EoS) the bar-mode instability only sets in at high rotation rates---when the star reaches 
$\beta_d \gtrsim 0.24-0.25$ \citep{2000ApJ...542..453S, 2007PhRvD..75d4023B},  
 where the rotation parameter is defined as $\beta = T/|W|$, with $T$  the kinetic and $W$ the gravitational potential energy. The threshold for instability is only marginally lower than  the Newtonian value, $\beta_d = 0.27$ \citep{1969efe..book.....C}, and it is not clear that remnants formed in a binary merger will get anywhere near this threshold.

However, differentially rotating stars may become dynamically unstable  at a (perhaps significantly) lower rotation rate. This was first shown by 
\citet{2001ApJ...550L.193C} and soon after confirmed by  \citet{2002MNRAS.334L..27S, 2003MNRAS.343..619S} and \citet{2003ApJ...595..352S}. 
These numerical simulations demonstrated that an instability may set in already for $\beta \simeq 0.01$. Given this low value, the mechanism is generally referred to as the low-T/W instability. %However it is worth to remark 
% that despite its name it can occur at any rotation rate.  
Since the original work, the low-T/W instability has been found in many physical scenarios; in  stellar core collapse   
 \citep{2005ApJ...625L.119O,  2007PhRvL..98z1101O, 2008A&A...490..231S, 2010ApJS..191..439K, 2016MNRAS.461L.112T, 2018MNRAS.475L..91T, 
2020MNRAS.tmpL..27S},  numerical evolutions of rapidly rotating cold neutron stars \citep{2006MNRAS.368.1429S, 2007CoPhC.177..288C, 2010CQGra..27k4104C} and simulations of binary post-merger remnants \citep{2019arXiv191004036D}.
It has also been demonstrated that
the interaction between magnetic field and differential rotation can decrease the amount of differential rotation and therefore suppress the low-T/W instability. This issue has been explored through both magnetohydrodynamical simulations \citep{2009ApJ...707.1610C, 2013PhRvD..88j4028F, 2014PhRvD..90j4014M} and equilibrium configurations of magnetised and differentially rotating stars \citep{2015MNRAS.450.4016F}.

The origin of the low-T/W instability is not yet well understood, although---as first proposed by \citet{2005ApJ...618L..37W}---there is strong evidence that it sets in when an oscillation mode enters co-rotation with the bulk motion, i.e.  
when the pattern speed of a given oscillation mode matches the local angular velocity of the star. The low-T/W instability can then be viewed as related to
local shear instabilities like the Papaloizou-Pringle instability in thick accretion discs \citep{1984MNRAS.208..721P}.  
In order to identify the instability, \citet{2006MNRAS.368.1429S} studied the behaviour of the canonical angular momentum \citep{1975ApJ...200..204F, 1978ApJ...221..937F, 1978ApJ...222..281F} in the region of  the co-rotation point, using both a linear method and hydrodynamical simulations.
The relation between the co-rotation point and the low-T/W instability was also confirmed by \citet{2015MNRAS.446..555P},  using time evolutions of the linearised equations to establish the instability onset and estimate the growth time for  sequences of differentially rotating polytropic stars (in Newtonian gravity). The results show that the $l=m=2$ f-mode becomes 
unstable as soon as it co-rotates with the star and the growth time tends to increase gradually as the mode moves deeper into the co-rotation band. Moreover,  the imaginary part of the f-mode can be described 
in terms of the stellar parameter and the mode pattern speed through an empirical formula.  In contrast, the r-mode does not appear  to suffer this instability, 
its pattern speed remains outside  the co-rotation region. It only approaches the boundary of the region for highly differentially rotating models \citep{PhysRevD.64.024003, 2015MNRAS.446..555P}.

The properties of modes suffering the low-T/W instability have been studied by \citet{2016PhRvD..94h4032S, 2017MNRAS.466..600Y} as an eigenvalue problem. In these studies 
the perturbation approach was restricted to the equatorial plane and compared to three-dimensional Newtonian hydrodynamical simulations. The results 
confirmed the relation between the growth time and the location of the co-rotation radius \citet{2015MNRAS.446..555P}, not only for the fundamental mode but also 
for pressure modes. Moreover, the results indicated a mode amplification between the co-rotation radius and the star's surface, suggesting that the unstable mode suffers an over-reflection at the co-rotation point. 
This analysis was later extended to stars described by equations of state with different stiffness \citep{2018PhRvD..98b4003S}. 

In the literature, differential rotation has mainly been described in terms of the j-constant rotation law \citep[see][]{2017LRR....20....7P}. However, more recent work (see \citet{2012A&A...541A.156G} and 
\citet{2016PhRvD..93d4056U, 2017PhRvD..96j3011U}) introduces new classes of rotation laws inspired by the differential rotation profile in merger remnants.  These  multi-parameter rotation laws  
 can be used to approximate the rotation properties of hypermassive neutron stars \citep{2015PhRvD..91f4027K, 2017PhRvD..96d3004H, 2017PhRvD..96d3019K, 2019arXiv191004036D}. 
The impact of these new rotation laws on the low-T/W instability and various oscillation modes has not been  explored so far. This is the problem we aim to address in the present work.  We set out to 
 consider these issues  by adapting the approach of \citet{2015MNRAS.446..555P}. This approach has been well tested and is known to produce stable long-term evolutions for perturbed neutron stars allowing for a precise extraction of the relevant oscillation mode features. Specifically, we study non-axisymmetric oscillations by evolving in time the linearised hydrodynamical equations in Newtonian gravity, for
 differentially rotating  stellar models described by one of the new rotation laws proposed by \citet{2017PhRvD..96j3011U}.  
With this set-up we can describe (albeit only at the qualitative level) the main features of the differential rotation observed in numerical simulations of binary neutron-star remnants.  
Due to the implicit mathematical form of the rotation law we have modified the original self-consistent method 
introduced by \citet{1986ApJS...61..479H}. We extract the mode frequencies and identify the unstable modes that drive the low-T/W instability 
by Fast Fourier Transformation (FFT) and monitor the canonical energy and angular momentum at the co-rotation point. The growth time of the unstable modes is determined by the kinetic energy time evolution. 

The results we present may provide  insight into the possible role of dynamical instabilities in neutron star merger remnants. Of course, realistic merger simulations are  much more complex than the linearised perturbation simulations we discuss here. Our focus is on the low-T/W instability and to what extent the dynamics changes when we consider merger-inspired rotation laws. This question is generally relevant, but our formulation requires a stable (at least on the timescale of the simulations) background configuration with respect to which we may define the perturbations. Such a configuration is unlikely to exist for the typical hypermassive remnants formed in mergers \citep{2017RPPh...80i6901B,bernuzzi}. And even if it does, there will not be a well-defined surface (as we assume) given that ejected matter will form a disk which may, in turn, exert a torque of the high-density matter. Our analysis does not allow us to consider nonlinear aspects, which be associated with complicated vortex dynamics \citep{2016PhRvD..94d4060K}, or relativistic features like the rotational frame dragging \citep{2015PhRvD..91f4027K}. The simple fact that our model is Newtonian also means that we cannot meaningfully consider realistic matter equations of state or, indeed, the key role played by heating and shocks. Our model is nowhere near realistic, but the results nevertheless provide a starting point for more detailed discussions of the problem, with some natural steps already taken by \citet{Xiaoyi2020}.

In Section~2 we summarize the formalism we have used, provide the relevant perturbation equations and describe the stellar models. Section~3 provides the results and Section~4 concludes the paper with a brief discussion of the implications. 

%%%%%%%%%%%%%%%%%%%%%%%%%%%%%%% SEC. %%%%%%%%%%%%%%%%%%%%%%%%%%%%%%%%%%%%%%%%%%%
\section{Formalism} \label{sec:Eqs}
%%%%%%%%%%%%%%%%%%%%%%%%%%%%%%%%%%%%%%%%%%%%%%%%%%%%%%%%%%%%%%%%%%%%%%%%%%%%%%%

\subsection{The Newtonian equations}
%%%%%%%%%%

In Newtonian gravity the equations required to study differentially rotating stars  are 
the Euler equation, the mass conservation equation and the Poisson equation for the gravitational potential: 
\begin{eqnarray}
\left( \frac{\partial}{\partial t} + \mtv{v} \cdot \nabla \right) \mtv{v}   & = & -  \nabla \left(  h +  \Phi  \right)
    \, ,  \label{eq:dvdt} \\
\frac{\partial \rho }{\partial t}  & = & - \nabla \cdot  \left(\rho \mtv{v} \right)    \, ,      \label{eq:drhodt} \\
\nabla^2 \Phi & = & 4 \pi G \, \rho \, , \label{eq:dPhi}
\end{eqnarray}
where $G$ is the gravitational constant. In these equations,
the scalar fields $\rho, h$ and $\Phi$ represent, respectively,  the mass
density, the specific enthalpy and the gravitational potential, while $\mtv{v}$
is the fluid velocity. 

The system of equations  is completed by an equation of state for the matter. In this work we are mainly interested in qualitative features,  
so it makes sense to consider a polytropic model;
\begin{equation}
P = k \rho ^{\gamma } \, , \label{eq:polEoS}
\end{equation}
where $k$ is a constant and the adiabatic index is given by
\begin{equation}
\gamma \equiv \frac{d \log P}{ d \log \rho}  = 1 + \frac{1}{n} \, , \label{eq:Gbdef}
\end{equation}
with $n$ is the polytropic index. All  results in this paper are obtained for  $n=1$. 
We do not expect that changing the stiffness of the model will modify at the qualitative level the properties we want to study.  
For instance the association between the instability and corotation points,  the relation between the growth time and the position 
of the corotation point. The impact of the EoS stiffness has been studied by \citet{2018PhRvD..98b4003S}.

In a barotropic fluid, pressure and enthalpy are related by 
\begin{equation}
h = \int  \frac{dP}{\rho}  \, , \label{eq:polEoSb}
\end{equation}
which for a polytropic model  leads to:
\begin{equation}
h = \frac{\gamma}{\gamma-1} \frac{P}{\rho} \, . \label{eq:polEoSc}
\end{equation}

%%%%%%%%%%%%%%%%%%%%%%%%%%%%%%% SEC. %%%%%%%%%%%%%%%%%%%%%%%%%%%%%%%%%%%%%%%%%%%
\subsection{Equilibrium solutions} \label{sec:ES}
%%%%%%%%%%%%%%%%%%%%%%%%%%%%%%% SEC. %%%%%%%%%%%%%%%%%%%%%%%%%%%%%%%%%%%%%%%%%%%

We consider sequences of axisymmetric and differentially rotating configurations determined from two different  rotation laws. 
Each  equilibrium model is a solution to  \citep{1986ApJS...61..479H}: 
\begin{equation}
\nabla \left(  h + \Phi  \right) - \frac{ \Omega^2  }{\! \! 2} \nabla \varpi ^2    = 0 \, , \label{eq:eq1}   
\end{equation}
which we can rewrite as 
\begin{equation}
\nabla \left(  h + \Phi  \right) = \frac{1}{2} \left( \Omega  \nabla j - j \nabla \Omega \right)  \, . \label{eq:1}   
\end{equation}
In order to obtain  equation (\ref{eq:1}) we have used the definition of the specific angular momentum $j=\Omega \, \varpi ^2$, 
where  $\Omega$ is the angular velocity  and  $\varpi = r \sin \theta$ is the radial distance from the rotation axis.  

\begin{figure}
%\begin{center}
\includegraphics[height=75mm]{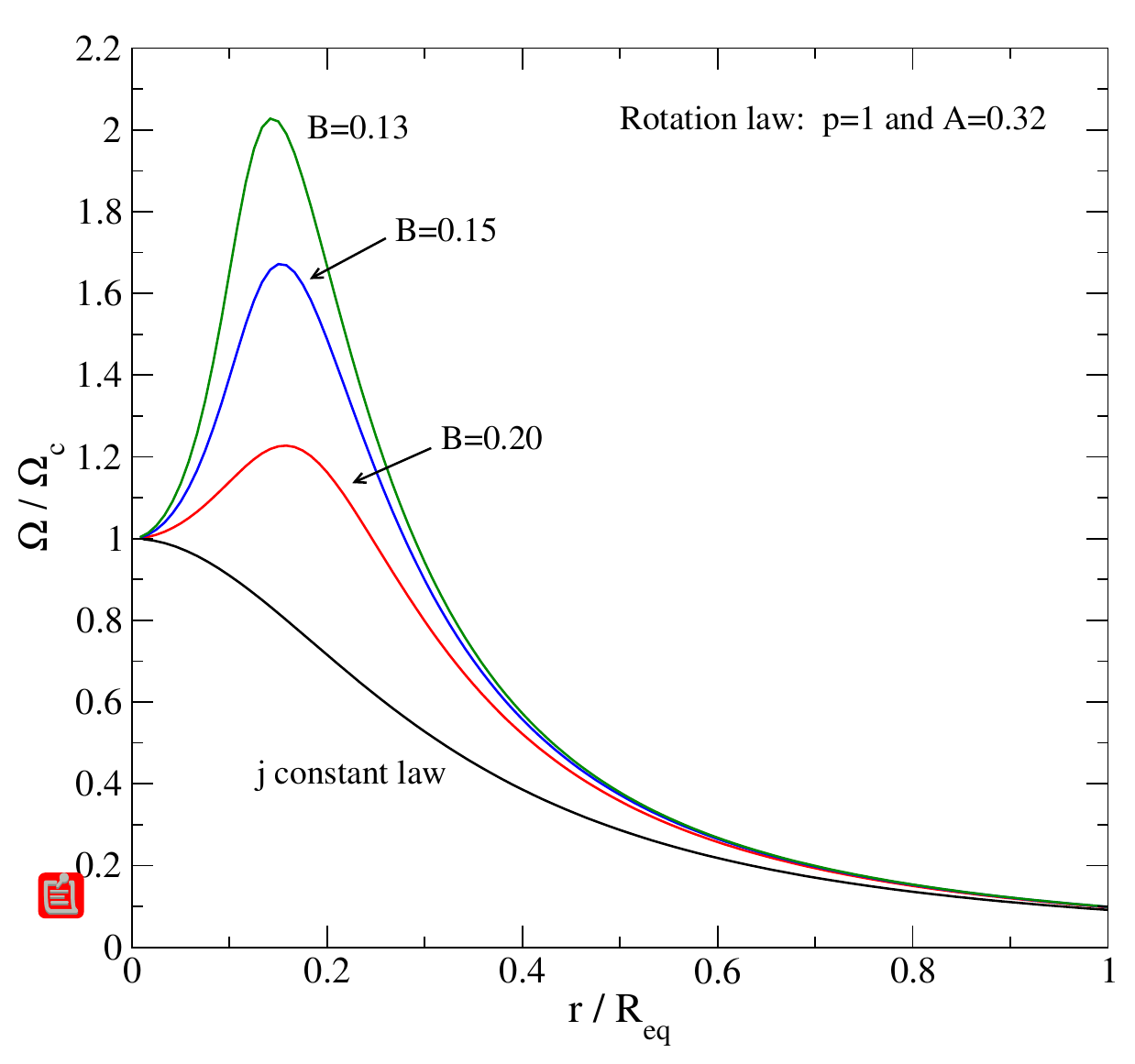} 
\includegraphics[height=75mm]{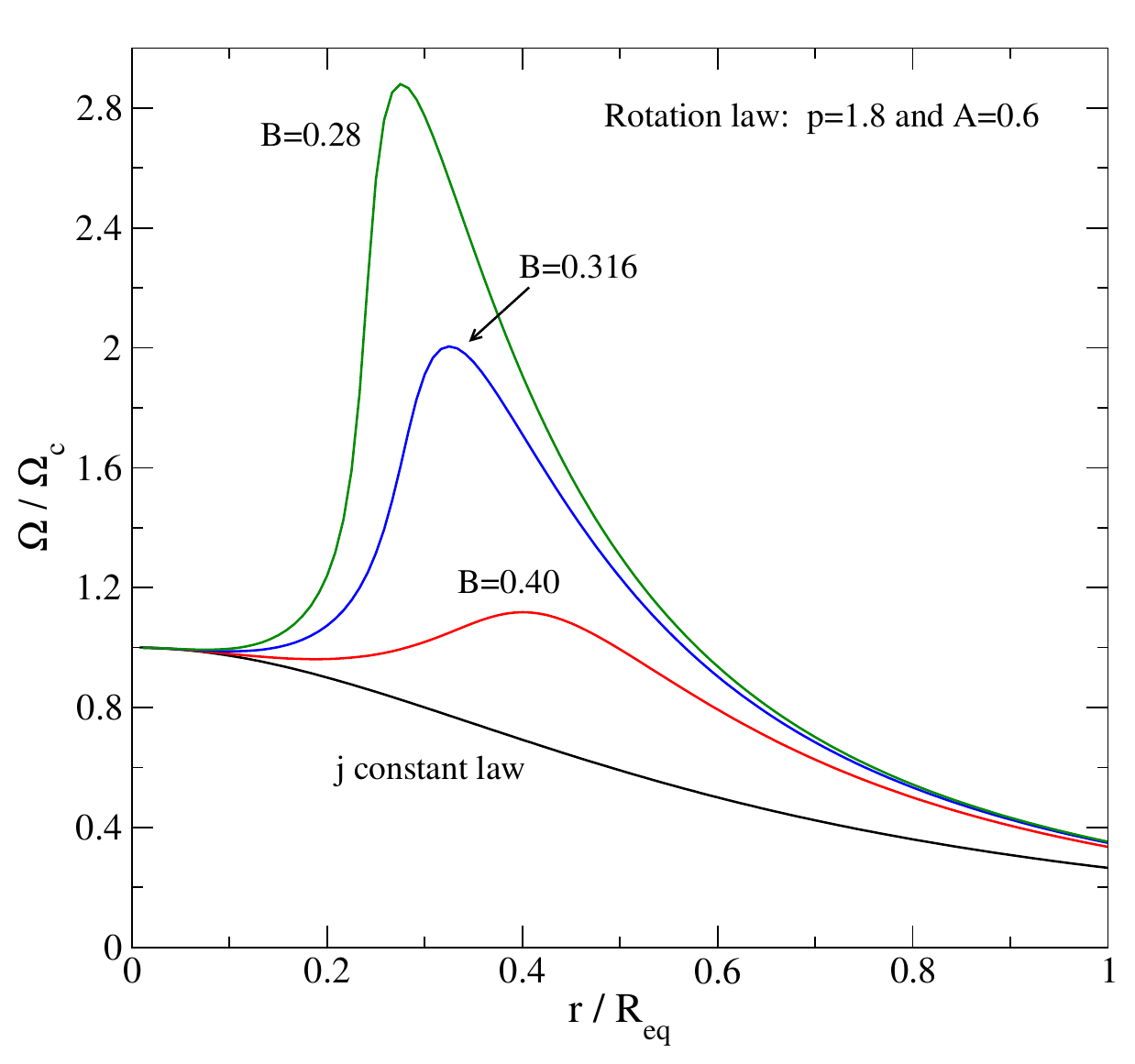} 
\caption{Rotation profiles obtained from equation (\ref{eq:Uryu-law}) for different stellar parameters. 
In the upper panel, we show three models having $p=1$, $\hat A=0.317$ and $B$, respectively given by $\hat B=0.20$, 0.15 and 0.13. 
In the lower panel, the star has $p=1.8$, $\hat A=0.6$ and the following three values of $\hat B$: 0.20, 0.15 and 0.13. 
For comparison we also show the solution corresponding to the j-constant rotation law. 
As pointed out in Section \ref{sec:star-model}, the rotation parameters $A$ and $B$ are given in dimensionless units, 
$\hat A = A/R_{eq}$ and $\hat B = B/R_{eq}$, where $R_{eq}$ is the equatorial radius. For clarity, we will continue to use $A$ and $B$ in the text and in the 
figure labels).} 
  \label{fig:prof}
%\end{center}
\end{figure}

The main new development in this paper relates to the implementation  of the   differential rotation law introduced by \citet{2017PhRvD..96j3011U},
\begin{equation}
\Omega =  \Omega_c \left[  1 + \left( \frac{j}{B^2 \Omega_c}  \right) ^{p} \, \right] \left(  1 - \frac{j}{A^2 \Omega_c}  \right) \label{eq:Uryu-law}   \, .
\end{equation}
Here, the quantity $\Omega_c$ denotes the angular velocity at the rotation axis and $p, A$ and $B$ are parameters which control the shape and the degree 
of differential rotation (see  figure \ref{fig:prof} for examples).  

If we introduce equation (\ref{eq:Uryu-law}) in the equilibrium equation (\ref{eq:1}) and integrate, we find the following expression:  
\begin{align}
& h + \Phi  \nn = C \\
& - \frac{\Omega_c}{2} j  \left[ 1 + \frac{1-p}{1+p} \left( \frac{j}{B^2 \Omega_c}  \right)^p  +    \frac{p}{2+p} \frac{1}{A^2 B^{2p}} \left( \frac{j}{\Omega_c} \right)^{p+1} \right] , \label{eq:4}  
\end{align}
where the integration constant $C$ is determined by imposing the required boundary condition at the pole. At this location,    
 the enthalpy and the specific angular momentum both vanish, leading to $C=\Phi_p$. Therefore, equation (\ref{eq:4}) becomes 
\begin{align}
& h = \Phi_p - \Phi   \nn \\
& - \frac{\Omega_c}{2} j  \left[ 1 + \frac{1-p}{1+p} \left( \frac{j}{B^2 \Omega_c}  \right)^p  +    \frac{p}{2+p} \frac{1}{A^2 B^{2p}} \left( \frac{j}{\Omega_c} \right)^{p+1} \right] . \label{eq:5}  
\end{align}

It is worth noting that, for $B \to \infty$ equation (\ref{eq:Uryu-law}) reduces to the well known j-constant law 
\begin{equation}
j = A^2 \left( \Omega_c -  \Omega \right)  \label{eq:j-law}   \,  , 
\end{equation}
which has an explicit expression in term of the star's angular velocity:
\begin{equation}
\Omega =    \frac{\Omega_c A^2}{A^2 + \varpi ^2 }   \label{eq:j-law2}   \,  ,
\end{equation}
a result that follows after using  the definition of the specific angular momentum.
%------------------------------FIG. 1------------------------------------------%
%------------------------------------------------------------------------------%

The main practical difference from implementations based on the j-constant law, relates to the fact that equation (\ref{eq:Uryu-law}) is an implicit  expression for  $\Omega$. 
This means that we have to modify the usual iterative method, described in \citet{1986ApJS...61..479H}.  We do this by the following steps: \\
1. We choose the polytropic index, $n=1$, the rotation parameters $(p,A,B)$ and the desired axis ratio. \\
2. From an initial guess density we solve the Poisson equation.  \\
3. We find the value of $\bar \Omega$ at the equator, where $\bar \Omega = \Omega / \Omega_c$, by solving the following equation with a root finder routine,
\begin{equation}
\bar \Omega =  \left(  1 +  \frac{\varpi^{2p}}{B^{2p}}  \bar \Omega ^p \right) \left(  1 - \frac{\varpi^2}{A^2 } \bar \Omega \right) \label{eq:alpha}   \, . 
\end{equation}
This relation is easily obtained from equation (\ref{eq:Uryu-law}). \\
4. We determine $\Omega_c$ by solving equation (\ref{eq:5}) at the equator, where (for a barotropic model) the enthalpy vanishes.\\ 
5. We solve equation (\ref{eq:5})  with a root finder routine to determine the enthalpy at each point. \\
6. We find the density profile, $\rho$, from the equation of state and repeat the iteration until the solution converges to the specified accuracy.

\begin{table}
%\begin{center}
\caption{\label{tab:back-models-0} This table lists key
quantities for two sequences of differentially rotating equilibrium
configurations. The stellar models are described by an  $n =1$ polytropic
equation of state and the j-constant 
rotation law. All quantities with the `hats' are expressed in dimensionless
units, where $G$ is the gravitational constant, $\rho_{\rm m}$ represents
the maximum mass density and $R_{\rm eq}$ is the equatorial radius. 
The first column gives the parameter $\hat A =  A/R_{eq}$ that
controls the degree of differential rotation. In the second, third and fourth 
columns, we provide, respectively, the ratio of polar to equatorial axes, 
the star's  mass $\hat M = M / (\rho_{\rm m} R_{\rm eq}^3)$ and the maximum pressure 
$  \hat p_{\rm m} = p_{\rm m}  / (G \rho_{\rm m}^2 R_{\rm eq}^2)$.
The fifth column displays the central angular velocity $\hat \Omega_c = \Omega_c  /  (G \rho_{\rm m})^{1/2} $ 
while the sixth column shows the angular momentum  $  \hat J = J / (G^{1/2} \rho_{\rm m}^{3/2} R_{\rm eq}^5)$.  
In the last column we provide the rotation parameter 
$ \beta = T/|W|$, the ratio between the rotational kinetic energy and gravitational
potential energy. The first row refers to the non-rotating model which is common to all the 
rotating sequences presented in this work. 
%All quantities are expressed in dimensionless
%units, where $G$ is the gravitational constant, $\rho_m$ represents
%the maximum mass density and $R_{eq}$ is the equatorial radius.
}
\begin{tabular}{c  c c c c c c }
\hline
% $A/R_{eq} $ &  $ R_p / R_{eq} $  &   $ M / (\rho_m R_{eq}^3)$   
 $\hat A $ &  $ R_p / R_{\rm eq} $  &  $  \hat M $   &  $ \hat  p_{\rm m}$
  & $  \hat  \Omega_c   $  & $ \hat  J $
  & $ \beta \times 10^{2}$ \\ %$\! T/ |W| \!\times \!10^{2}$ \\
\hline
%A.     Rax		    M		    p_max      Om_c            J             T/W					
       &  1.0            &    1.273     &    0.637   & 0.000     &    0.0000   &   0.00      \\  \\
       
 0.32 &  0.9          &    1.277    &     0.597   &  1.432     &   0.1147   &   1.98    \\
 0.32 &  0.7          &    1.295    &     0.510   &  2.630     &   0.2204   &   6.44    \\
 0.32 &  0.5          &    1.241    &     0.419   &  3.447     &   0.2885   &   11.31   \\
 0.32 &  0.3          &    1.143    &     0.338   &  3.909     &   0.3167   &  15.54    \\
 0.32 &  0.1          &    1.066    &     0.293   &  4.046     &   0.3166   &  17.64   \\ \\

 0.60 &  0.9            &    1.211   &     0.580  &  0.794     &   0.1268   &   2.41     \\
 0.60 &  0.7            &    1.097   &     0.461  &  1.396     &   0.2087   &   8.05     \\
 0.60 &  0.5            &    1.034   &     0.337  &  1.844     &   0.2748   &  14.91    \\
 0.60 &  0.3            &    0.930   &     0.230  &   2.049    &   0.3044   &  21.75    \\
 0.60 &  0.1            &    0.823   &     0.183  &  1.954     &   0.2829   &  24.13     \\
\hline   
\end{tabular}
%\end{center}
\end{table}
%------------------------------------------------------------------------------%

%------------------------------TAB. 1------------------------------------------%
\begin{table}
%\begin{center}
\caption{\label{tab:back-models-1} 
Same physical quantities as in Table \ref{tab:back-models-0}, but for  models 
with $p=1$ and $\hat A=0.32$. The first column now reports the parameter $B$.}
\begin{tabular}{c  c c c c c c  }
\hline
 $ \hat B $ &  $ R_\p / R_{\rm eq} $  &  $ \hat M $ &  $ \hat p_{\rm m} $ 
  & $ \hat \Omega_c  $  &  $  \hat J $ &  $  \beta \times 10^{2}$ \\
\hline
%B.     Rax		    M		       Om_c           T/W					
%       &  1.0   &    1.273      &     0.000     &   0.00      \\ \\
 0.20 &  0.9  &    1.296   & 0.601 &   1.046     & 0.1072  &   1.89     \\
 0.20 &  0.7  &    1.346   & 0.524 &   1.939     & 0.2119  &    6.01   \\
 0.20 &  0.5  &    1.285   & 0.443 &   2.524     & 0.2723  &  10.23    \\
 0.20 &  0.3  &    1.198   & 0.373 &   2.803     & 0.2988  & 13.59       \\
 0.20 &  0.1  &    1.161   & 0.334 &   2.962     & 0.3033  & 14.64     \\
\\
 0.15 &  0.9  &    1.311   & 0.605  &   0.909     & 0.1003 &   1.74    \\
 0.15 &  0.7  &    1.355   & 0.535  &   1.680     & 0.1961 &   5.48  \\
 0.15 &  0.5  &    1.302   & 0.463  &   2.192     & 0.2527 &   9.22  \\
 0.15 &  0.3  &    1.232   & 0.400  &   2.516     & 0.2813 &  12.11   \\
 0.15 &  0.1  &    1.184   & 0.366  &   2.662     & 0.2902 &  13.53  \\
\\
 0.13 &  0.9  &   1.319    & 0.607  &  0.838      & 0.0957  &  1.64 \\
 0.13 &  0.7  &   1.358    & 0.542  &  1.545      & 0.1855  &  5.14 \\
 0.13 &  0.5  &   1.312    & 0.475  &  2.022      & 0.2399  &  8.58 \\
 0.13 &  0.3  &   1.251    & 0.417  &  2.332      & 0.2697  & 11.22 \\
 0.13 &  0.1  &   1.209    & 0.385  &  2.480      & 0.2804  & 12.52 \\
\hline   
\end{tabular}
%\end{center}
\end{table}
%------------------------------------------------------------------------------%

%------------------------------TAB. 2------------------------------------------%
\begin{table}
%\begin{center}
\caption{\label{tab:back-models-2}  Same physical quantities as in Table \ref{tab:back-models-1}, but for  models 
with $p=1.8$ and $\hat A=0.6$.}
\begin{tabular}{c  c c c c c c c  c }
\hline
$ \hat B $ &  $ R_\p / R_{\rm eq} $  &  $ \hat M $ &  $ \hat p_{\rm m} $ 
  & $ \hat \Omega_c  $  &  $  \hat J $ &  $  \beta \times 10^{2}$ \\
\hline
%B.     Rax		    M		       Om_c           T/W					
 0.40 &  0.9            &      1.201     & 0.577 &      0.546    & 0.1308  &   2.64  \\
 0.40 &  0.7            &      1.068     & 0.452 &      0.949    & 0.2106  &   8.82 \\
 0.40 &  0.5            &      1.008     & 0.324 &      1.240    & 0.2787  & 16.27  \\
 0.40 &  0.3            &      0.954     & 0.219 &      1.366    & 0.3287  &  22.82  \\
 0.40 &  0.1            &      0.819     & 0.179 &      1.253    & 0.2833  &  24.02  \\
\\
 0.316 &  0.9          &     1.227     & 0.583  &   0.418      & 0.1275 &   2.59    \\
 0.316 &  0.7          &     1.165     & 0.470  &   0.743      & 0.2248 &   8.51  \\
 0.316 &  0.5          &     1.205     & 0.355  &   1.003      & 0.3349 &  15.01  \\
 0.316 &  0.3          &     1.019     & 0.261  &   1.046      & 0.3213 &  19.80   \\
 0.316 &  0.1          &     0.908     & 0.224  &   1.013      & 0.2920 &  20.92  \\
\\
 0.28 &  0.9           &   1.248  & 5.8831 &  0.359    & 0.1222  & 2.46     \\
 0.28 &  0.7           &   1.241  & 4.8543 &  0.650    & 0.2293  & 7.94     \\
 0.28 &  0.5           &   1.228  & 3.8032 &  0.856    & 0.3151  & 13.66   \\
 0.28 &  0.3           &   1.073  & 2.9384 &  0.918    & 0.3145  & 17.71   \\
 0.28 &  0.1    	    &    0.974 & 2.5358 &  0.917    & 0.2953  & 18.81   \\
\hline   
\end{tabular}
%\end{center}
\end{table}
%------------------------------------------------------------------------------%

%%%%%%%%%%%%%%%%%%%%%%%%%%%%%%% SEC. %%%%%%%%%%%%%%%%%%%%%%%%%%%%%%%%%%%%%%%%%%%
\subsection{Stellar models} \label{sec:star-model}
%%%%%%%%%%%%%%%%%%%%%%%%%%%%%%%%%%%%%%%%%%%%%%%%%%%%%%%%%%%%%%%%%%%%%%%%%%%%%%%%

With the method outlined in Section \ref{sec:ES} we can construct  stellar models to explore the effects of the rotation laws on the mode frequency and low T/W instability.   We need to fix three parameters of the rotation law (\ref{eq:Uryu-law}): the index $p$ and the two 
constants $A$ and $B$ which control the degree of differential rotation. The rotation rate at the center, $\Omega_c$, is determined 
indirectly via the iterative method we outlined above. The parameter $A$, which is also present  in the j-constant rotation law, 
controls the ratio between the equatorial and axial rotation rate. The parameter $B$ and the index $p$ in equation (\ref{eq:Uryu-law}) mainly affect  
the shape of the rotation profile and the position of a maximum away from the rotation axis (see figure \ref{fig:prof}).

We consider, for our explorative work, two sets of solutions with rotation properties similar to the remnants of binary neutron-star mergers  from nonlinear numerical evolutions \citep{2015PhRvD..91f4027K, 2017PhRvD..96d3004H, 2017PhRvD..96d3019K, 2019arXiv191004036D}. 
   In these two sets of models we first choose and keep constant the parameters $p$ and $A$ while the parameter $B$ is varied to change the position and magnitude of the peak. 
For the first set of  models we choose $p=1$, $A=0.316$ and three different values of $B$, namely $B=0.2$, 0.15 and 0.13 (note that the rotation parameters $A$ and $B$ are given in dimensionless units, $\hat A = A/R_{eq}$ and $\hat B = B/R_{eq}$, where $R_{eq}$ is the equatorial radius, but for clarity, we will continue to use $A$ and $B$ in the text). 
For these values we construct three sequences of differentially rotating stars, from the non-rotating models up to the extreme case of an axis ratio $R_p / R_{\rm eq}= 0.05$, where $R_p$ and $R_{eq}$ are, respectively, the stellar radius 
at the pole and at the equator. 
Figure \ref{fig:prof} shows the rotation profile normalised to the  central angular velocity, the effects of the parameter B  on the shape of the rotation profile is clear. The maximum of $\Omega$ increases for smaller $B$. 

The second family of models has $p=1.8$, $A=0.6$ and the three values  $B=0.4$, 0.316, 0.28.
The rotation profiles for  these models are shown in figure \ref{fig:prof} together with the j-constant law for $A=0.316$ and $0.6$. 
By varying the rotational parameters we have constructed models with different maximum angular velocity, $\Omega_{\rm max}$, and degree of differential rotation. For instance,
figure \ref{fig:prof} shows that the quantity $\Omega_{\rm max} / \Omega_{\rm c}$ varies between 1 and 2.8, while $\Omega_{\rm eq} / \Omega_{\rm c}$ 
is practically equal for all models with $p=1$. Meanwhile, for the case with $p=1.8$ the j-constant law leads to a slightly smaller value of $\Omega_{\rm eq} / \Omega_{\rm c}$ compared to the other three models. 
The main quantities of the models we use are reported, in dimensionless units, in Tables \ref{tab:back-models-0}, \ref{tab:back-models-1} and \ref{tab:back-models-2}. 
These dimensionless units are defined in terms of the gravitational constant $G$, maximum mass density $\rho_{\rm m}$ and equatorial radius $R_{\rm eq}$ 
        (see caption of Table \ref{tab:back-models-0}). 
        For the polytropic $\gamma=2$ EoS, one can construct, from the results shown in Tables \ref{tab:back-models-0}, \ref{tab:back-models-1} and \ref{tab:back-models-2}, different sequences of rotating stars, e.g. with constant mass or constant angular momentum.
        For instance, by specifying the mass of the star M and the EoS parameters $k$ and $\gamma $, we can obtain the equatorial radius and the maximum mass density in physical units from the following expressions:

\begin{eqnarray}
&& R_{\rm eq} = \left[  \frac{1}{G} \frac{k}{\hat k} \left( \frac{M}{\hat M} \right)^{\gamma - 2} \right] ^{1/(3 \gamma - 4 )}  \label{eq:Rp}\, \\
&& \rho_{\rm m} = M \hat M^{-1} R_{\rm eq} ^{-3}  \label{eq:rhop} \, .
\end{eqnarray}
where for a polytropic EoS $\hat k$ is equal to the maximum dimensionless pressure $\hat p_{\rm m}$ , and the `hats' are the dimensionless quantities shown in Table \ref{tab:back-models-0}, \ref{tab:back-models-1} and \ref{tab:back-models-2}.
As an example, we outline here a method to construct a sequence of rotating models  with constant mass. We consider for simplicity models described by the j constant rotation law with $A = 0.32 $, an extension  to all the other cases is straightforward. \\
1. First of all, we choose the physical stellar mass, e.g. $M=2 M_{\odot}$ and determine the properties of the nonrotating model. From the first row of Table \ref{tab:back-models-0} we read 
$\hat M$ = 1.273 and $\hat{p}_m = 0.637$  and we determine  $R_{eq}$ and $\rho_{m}$ by using equations (\ref{eq:Rp}) and (\ref{eq:rhop}).  \\
2. For a rotating star, we select the model with $\beta = 0.0198 $  (second row of Table \ref{tab:back-models-0}) and we read 
the dimensionless quantities $\hat M = 1.277$,  $\hat{p}_m= 0.597$, $\hat \Omega_c = 1.432$ and $\hat{J} = 0.1147$.   
From equations (\ref{eq:Rp}) and (\ref{eq:rhop}) we determine $R_{eq}$ and $\rho_{m}$, which we can use to calculate all the other quantities in physical units. 
For instance $\Omega_c = \hat \Omega_c \times \sqrt{G \rho_m}$ and $J = \hat J \times G^{1/2} \rho_m^{3/2} R_{eq}^5 $. 

%A rotating model. For instance the model with A_hat = 0.32 and beta = 0.0198 (second line of table 1). We take the values M_hat = 1.277, pm_hat = 0.597, Omega_c_hat = 1.432 and J_hat = 0.1147.   
%From equation (16)-(17) we determine the new Req and rho_max. With these values I determine the physical Omega_c = Omega_c_hat x sqrt(G rho_max) and the physical J = J_hat x ( G^(1/2) rho_max^(3/2) Req^5 ). 

%The same method can be used for any other model. 

%%%%%%%%%%%%%%%%%%%%%%%%%%%%%%% SEC. %%%%%%%%%%%%%%%%%%%%%%%%%%%%%%%%%%%%%%%%%%%
\subsection{Perturbation equations} \label{sec:pert-eqs}
%%%%%%%%%%%%%%%%%%%%%%%%%%%%%%%%%%%%%%%%%%%%%%%%%%%%%%%%%%%%%%%%%%%%%%%%%%%%%%%%

Turning to the dynamical aspects of the problem, we now briefly review our approach to the oscillation modes and the properties of the low-T/W instability (more details can be found in \citet{2015MNRAS.446..555P}). 

We study  non-axisymmetric oscillations of differentially rotating stars by using as dynamical variables 
 the enthalpy $\delta h$ and the velocity perturbation $\delta \mtv{v}$, which,  for an inertial observer and in spherical coordinates $[r,\theta,\phi]$, 
 obey the linearised equations 
\begin{eqnarray}
\left( \frac{\partial}{\partial t} + \Omega \frac{\partial}{\partial \phi} \right) \delta \mtv{v} & = & 
                     - \nabla \delta h - 2 \mtb{\Omega} \times \delta \mtv{v}                
                      \nn \\ 
                     &&
                     - \left( \delta \mtv{v} \cdot \nabla \Omega \right) r \sin\theta \, \hat{e}_{\phi}
                       \, ,                       \label{eq:dfdt}  \\
\left( \frac{\partial}{\partial t} + \Omega \frac{\partial}{\partial \phi} \right) \delta h  & = & -  \frac{\partial h }{\partial \rho} \, \nabla \cdot  \left( \rho \delta \mtv{v}  \right) \, ,      \label{eq:dPdt} 
\end{eqnarray}
where $\hat{e}_{\phi}$ is the $\phi$-component of the orthonormal basis.  
We use the Cowling approximation, i.e.  neglect the gravitational potential perturbation $\delta \Phi$ during the numerical evolution. 
The benefit of this is that we do not have to solve the linearised Poisson equation (an elliptic equation), speeding up the evolution.  The drawback is that we  lose accuracy in the calculation of the f-mode frequencies   \citep{2003MNRAS.343..175K}, but this is not a major issue in this exploratory  study. 
Our current results would anyway change when we  consider the problem in General Relativity \citep{Xiaoyi2020}.
The accuracy of the Cowling approximation for this kind of problem has been already discussed in \citet{2015MNRAS.446..555P}, in the context of the full set of Newtonian 
perturbation equations.

For axisymmetric stars, we can simplify the problem by using a Fourier expansion in the $\phi$ coordinate 
of equations (\ref{eq:dfdt})-(\ref{eq:dPdt})~\citep{1980MNRAS.190...43P}. 
This way we end up with a two-dimensional problem, because the perturbation variables depend only on the spatial coordinates $(r,\theta)$ and the azimuthal index $m$. 
More specifically, any perturbation variable can be written in a similar way to the enthalpy:
\begin{equation}
\!  \delta h \left( t,r,\theta,\phi \right) = \! \!  \! \sum_{m=0}^{m=\infty}
               \!  \left[ \delta h_{m}^{+} \left( t,r,\theta\right)
               \cos m \phi  + \!  \delta h_{m}^{-} \left(
               t,r,\theta\right) \sin m \phi \right] \, .
               \label{eq:dPexp}
\end{equation}
With this approach we  numerically evolve, for any chosen $m$, a system of eight  partial differential
equations for $\left( \delta \mtv{v}^{\pm}, \delta
h^{\pm} \right)$.

In Section \ref{sec:Enc} we will introduce the canonical energy and canonical angular momentum, the main diagnostics  we  use to 
identify  the position where the instability develops. In order to determine these two quantities we need  the  
Lagrangian displacement vector $\pmb{\xi}$~\citep{1978ApJ...221..937F}. Therefore, we have to solve (at each time step) 
\begin{equation}
\delta \mtv{v} = \frac{\partial \pmb{\xi}}{\partial t} + \mtv{v} \cdot \nabla \pmb{\xi} -  \pmb{\xi} \cdot \nabla \mtv{v}  \label{eq:xi} \,  .
\end{equation}

%------------------------------FIG. 2------------------------------------------%
\begin{figure*}
\begin{center}
\includegraphics[height=72mm]{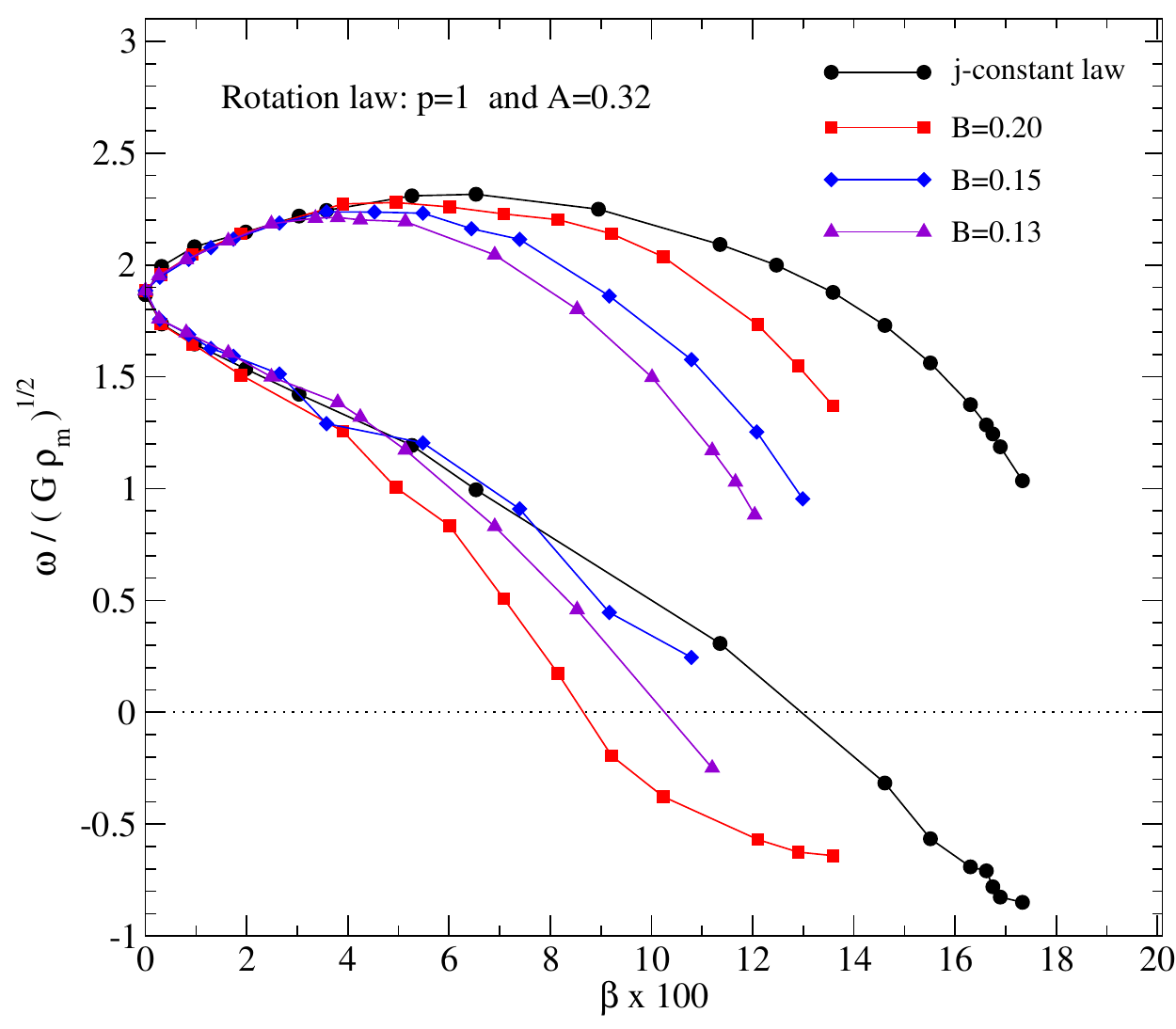} \hspace{0.5cm} 
\includegraphics[height=72mm]{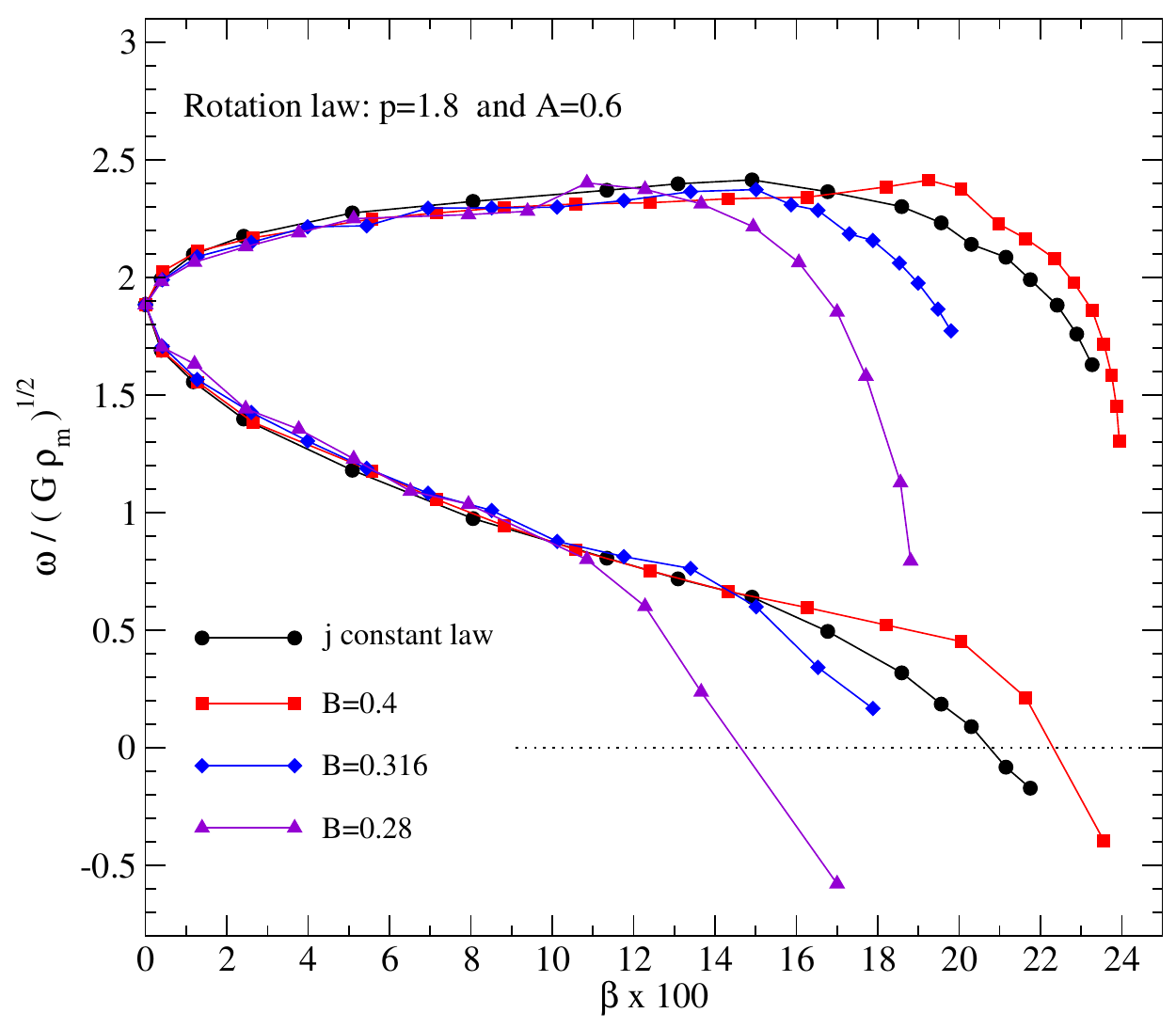} 
\caption{Variation of the $\rm^{2}f$ mode frequency with $\beta = T/|W|$ for the  models shown in figure \ref{fig:prof}. 
The mode frequencies of the three models with $p=1$ and $A=0.32$  are shown in the the left-hand panel (see legend), while those for the three models with $p=1.8$ and $A=0.6$ are shown 
in the right-hand panel. The horizontal dotted line denotes the neutral point of the CFS instability (see Section \ref{sec:f-mode}) \label{fig:f2}}
\end{center}
\end{figure*}
%------------------------------------------------------------------------------%

%%%%%%%%%%%%%%%%%%%%%%%%%%%%%%% SEC. %%%%%%%%%%%%%%%%%%%%%%%%%%%%%%%%%%%%%%%%%%%
%\subsubsection{Boundary conditions} \label{sec:BC}
%%%%%%%%%%%%%%%%%%%%%%%%%%%%%%%%%%%%%%%%%%%%%%%%%%%%%%%%%%%%%%%%%%%%%%%%%%%%%%%%

Finally, we need to impose the relevant boundary conditions. 
At the origin, $r=0$, and on the rotational axis, $\theta = 0$, 
the solutions to equations~(\ref{eq:dfdt})-(\ref{eq:dPdt}) must remain regular. 
For non-axisymmetric oscillations with $m \geq 2$, this condition is realised by imposing 
\begin{equation}
\delta h = \delta \rho = 0 \, ,  \quad \textrm{and} \quad \delta \mtv{v} = 0 \, .
\end{equation}

At the stellar surface we require that the Lagrangian
perturbation of the enthalpy vanishes, i.e.,
\begin{equation}
\Delta h = \delta h + \pmb{\xi}  \cdot \nabla h = 0  \, . \label{eq:DP-bc}
\end{equation}
For a barotropic model, this is equivalent to the vanishing of the Lagrangian variation in pressure.
We satisfy this boundary condition by imposing 
\begin{equation}
\delta h  =  - \pmb{\xi}  \cdot \nabla h \, ,\label{eq:DP-bc2}
\end{equation}
at the surface.
All  other variables are extrapolated at the surface
grid point at each time step.

Finally, at the equator, $\theta = \pi/2$, the 
perturbation variables divide into two classes with opposite reflection
symmetry. In the first class,  the  variables $\delta h^+  , \delta v_r^+, \delta v_{\phi
}^+$ are all even under reflection with respect to the equatorial
plane, while $\delta v_{\theta}^- $ is odd.  
In contrast, for the second
class $\delta h^- , \delta v_r^-, \delta v_{\phi}^-$ are odd and $\delta v_{\theta}^+ $ is even.

%%%%%%%%%%%%%%%%%%%%%%%%%%%%%%%%%%%%%%%%%
\subsection{Canonical energy and angular momentum} \label{sec:Enc}
%%%%%%%%%%%%%%%%%%%%%%%%%%%%%%%%%%%%

At the linear perturbation level, 
the development of non-axisymmetric instabilities in rotating bodies can be monitored in terms of the 
canonical energy and angular momentum~\citep{1975ApJ...200..204F, 1978ApJ...221..937F, 1978ApJ...222..281F}.  
In an inviscid star, an unstable mode does not violate the energy and angular momentum conservation 
laws. This means that such a mode may only grow if both the canonical energy and angular momentum vanish.  
As explained in  \citet{2015MNRAS.446..555P}, our numerical approach (based on time evolving the perturbation equations) is not accurate 
enough to directly monitor these two conditions, mainly because of the presence of other oscillation modes in the evolved data, the general issue of keeling track of a growing unstable mode  and  the 
numerical viscosity which  contaminates the eigenfunction extraction. 
However, as shown by   \citet{2006MNRAS.368.1429S} and confirmed by  \citet{2015MNRAS.446..555P} 
the canonical energy and angular momentum still helps identify the location inside the star where the low-T/W instability develops. 

The canonical energy  is given by~\citep{1978ApJ...221..937F}: 
\begin{align}
 E_{c}  & = \frac{1}{2} \int d \mtv{r}  \left[  \rho | \partial_{t} \xi_{i} |^{2} - \rho | v^{j} \nabla_{j} \xi_{i}  |^{2} 
+ \rho \, \xi^{i} \xi^{j \ast} \nabla_{i} \nabla_{j} \left( h + \Phi \right)  \right. \nn \\
& \left.+ \frac{\partial h}{\partial \rho} \, | \delta \rho | ^2 
- \frac{1}{4\pi G} | \nabla_{i} \delta \Phi | ^{2}  \right] \, , \label{eq:Ec} 
\end{align}
while the canonical angular momentum follows from: 
\begin{equation}
J_{c} = - \mathrm{Re} \int d \mtv{r}  \rho \, \partial_{\phi} \xi^{i \ast} \left( \partial_{t} \xi_{i} +  v^{j} \nabla_{j} \xi_{i}  \right)  . \label{eq:Jc} 
\end{equation}
The integrals are calculated over the star's volume, and  
 Re denotes the real part.  Note that these expressions are given in a coordinate basis, not the orthonormal basis used elsewhere in the paper.

%%%%%%%%%%%%%%%%%%%%%%%%  SEC: Code  %%%%%%%%%%%%%%%%%%%%%%%%%%%%%%%
\subsection{Code description and tests} \label{sec:code}
%%%%%%%%%%%%%%%%%%%%%%%%%%%%%%%%%%%%%%%%%%%%%%%%%%%%%%%%%

We use the  code developed by \citet{2015MNRAS.446..555P} to study the time evolution of the linearised equations. 
The numerical grid is two-dimensional in the coordinates $(r,\theta)$ which lie in the range: $0 \leq r \leq R(\theta) $ and $ 0 \leq \theta
\leq \pi / 2$.  With a new definition of the radial coordinates $x = r/R(\theta)$ we adapt the grid to the star even 
when the fluid is highly deformed by rotation.  The perturbation equations are 
discretized on the grid and updated in time with a Mac-Cormack
algorithm. Finally, the numerical simulations are stabilised from high
frequency noise with the implementation of a fourth order
Kreiss-Oliger numerical dissipation $\veps_{\rm D} D_{4} \pmb \xi $, with $\veps_{\rm D} \approx 0.01$. More technical details on the numerical implementation can be found in~\citet{2015MNRAS.446..555P} and \citet{2009MNRAS.394..730P, 2009MNRAS.396..951P}.

Most of the results discussed in this paper were obtained using a $48\times90$ grid to cover the $\theta$ and $r$ coordinates, respectively. 
In order to test the accuracy of our instability growth time extraction we evolved some models on a 
$96\times180$ grid. This showed that the numerical error in the key quantities was significantly less than 1\%, which means that the conclusions we draw from the results should be reliable.

%%%%%%%%%%%%%%%%%%%%%%%%  SEC: Code  %%%%%%%%%%%%%%%%%%%%%%%%%%%%%%%
\section{Results} \label{sec:result}
%%%%%%%%%%%%%%%%%%%%%%%%%%%%%%%%%%%%%%%%%%%%%%%%%%%%%%%%%

In our previous work we studied the relation between  co-rotating modes and the low-T/W instability for differentially rotating stars \citep{2015MNRAS.446..555P}. 
The rotation profile was modelled by the j-constant rotation law. As suggested by \cite{2005ApJ...618L..37W}  we found---within the accuracy of the numerical framework---that the instability 
  sets in when the f-mode enters the co-rotation region, i.e. whenever there is a point where the pattern speed of the mode matches the local rotation velocity of the star. 
  This means that  
  \begin{equation}
 \sigma = \Omega (\varpi_c) \,  , \label{eq:cor}
\end{equation}
 where $\sigma = \omega / m$ is the pattern speed of the mode and $\varpi_c$ is the co-rotation point. 

In this work, we consider the impact of the  rotation law from (\ref{eq:Uryu-law}) on the mode frequencies and the instability growth time. This question is interesting because the new rotation laws relates more directly to the dynamics of neutron star merger remnants and, as is evident from figure~\ref{fig:prof}, a given mode may have two co-rotation points inside the star. If the mode entering co-rotation is a requirement for the instability to be triggered, then does the presence of additional co-rotation points impact on this?

First of all, we construct for each stellar model a sequence of differentially rotating stars---from the non rotating 
case up to very rapidly rotating configurations. For each member of the rotating sequence we evolve in time the linearised equations 
and extract the mode frequencies via a FFT on the evolved quantities. 
For unstable models, we identify the mode frequency---presumably related to the instability being triggered---and determine 
the presence of a associated co-rotation radius.  
To support the mode identification we also extract the 2D eigenfunctions by using 
a code developed by \citet{2004MNRAS.352..1089S}. 
Finally,  we monitor both the canonical energy density and the angular momentum density to check that these quantities have the expected growth on both sides 
 of the co-rotation radius \citep{2006MNRAS.368.1429S, 2015MNRAS.446..555P}. 

To establish the growth time $\tau$ of the unstable mode, we focus on the kinetic energy,
\begin{equation}
E_k = \frac{1}{2} \int d \mtv{r} \,  \rho \, \delta \!\mtv{v}^2 \, ,
\end{equation}
and assume that 
\begin{equation}
E_k \sim e^{ 2  \omega_I t } \, , \qquad  \textrm{where} \quad \omega_I =   \frac{2 {\rm \pi}}{\tau} \, .  \label{eq:omI}
\end{equation}
When the instability sets in the kinetic energy starts to grow exponentially and we determine $\omega_I$ from a linear fit of 
the time evolved  energy \citep[see][for more details]{2015MNRAS.446..555P}.

%------------------------------FIG. 3------------------------------------------%
\begin{figure}
%\begin{center}
%\includegraphics[height=69mm]{Fig/f2-ur-a0_6-bvar.pdf} 
\includegraphics[height=69mm]{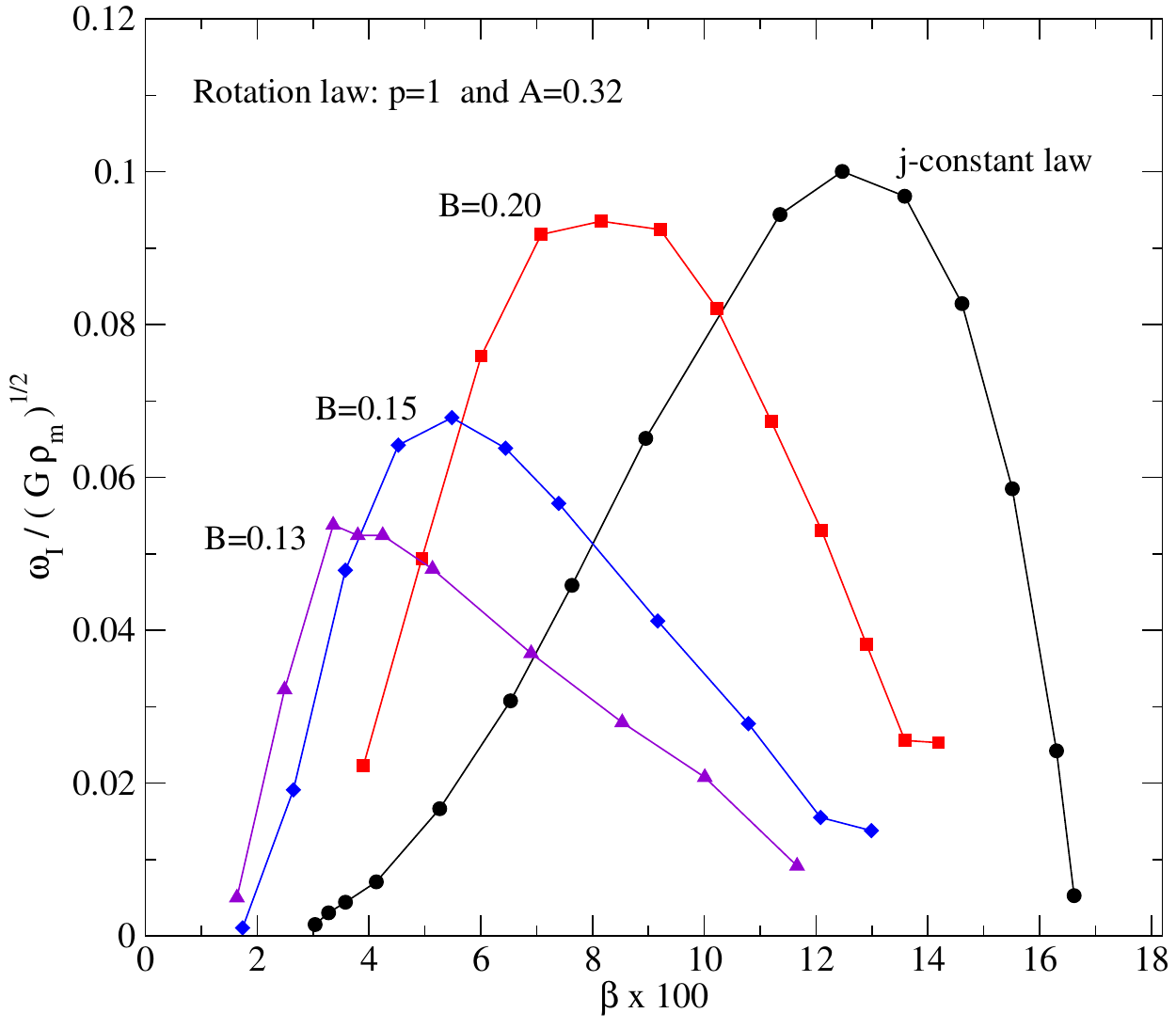} 
\caption{This figure shows, for  rotating sequences with $p=1$ and $A=0.317$ and respectively $B=0.2, 0.15$ and 0.13,  how the quantity  $\omega_I = 2\pi / \tau$ varies with the rotation parameter $\beta$. For comparison we also show the results for the j-constant model with $A=0.317$. The results suggest that, while the instability sets in for lower values of $\beta$ as $B$ decreases, the fastest growth rate is obtained for the j-constant model.
\label{fig:omI_A}}
%\end{center}
\end{figure}
%------------------------------------------------------------------------------%

%%%%%%%%%%%%%%%%%%%%%%%%  SEC: Code  %%%%%%%%%%%%%%%%%%%%%%%%%%%%%%%
\subsection{The f-mode} \label{sec:f-mode}
%%%%%%%%%%%%%%%%%%%%%%%%%%%%%%%%%%%%%%%%%%%%%%%%%%%%%%%%%

Before we consider the low-T/W instability we explore the effects of the new differential laws on the f-mode frequencies. 
We focus on the quadrupole mode  ($l=m=2$), which is the most important for gravitational wave emission. 
Figure \ref{fig:f2} shows the $^2$f-mode frequencies, measured in the inertial reference frame,  for the eight sequences of models  illustrated in figure \ref{fig:prof}.  In the left-hand panel, all  stars have $p=1$ and $A=0.317$ but different values of $B$ (see legend of figure \ref{fig:f2}). The  
sequence associated with the j-constant  law has the lowest level of differential rotation and is simply obtained 
by setting $B=100$.   Similarly,  we show in the right-hand panel of figure \ref{fig:f2} the f-mode frequencies for models with $p=1.8$ and $A=0.6$ but different $B$. 
The fastest spinning stellar models of these sequences assume a ``toroidal-like'' configuration 
with a small axis ratio and a mass density whose maximum is shifted away from the rotation axis. 

The $^2 \rm f $ mode is split by rotation, as expected, into two branches, which are  prograde and retrograde with respect to the star rotation. 
In both panels of figure~\ref{fig:f2}, we notice a similar behaviour of the mode frequencies with regard to 
$ \beta \equiv T/|W|$, the ratio between the rotational kinetic energy and gravitational
potential energy. 
For lower rotation rates the splitting of the  $^2 \rm f $ mode seems largely independent of the parameter $B$. In fact, 
figure  \ref{fig:f2} suggests that models with $p=1$ and $A=0.317$ have very similar f-mode frequencies when  $\beta \lesssim 0.4$. 
Models with $p=1.8$ and $A=0.6$ shows the same behaviour for  $\beta \lesssim 0.10-0.11$. The main lesson may be that, up to these rotation rates the f-mode frequency is adequately described by the usual j-constant prescription. 
For faster rotation rates, the parameter $B$ strongly affects the mode frequencies but apparently not in a unique way. 
For models with $p=1$ and $A=0.317$ we see that the mode frequencies decrease comparing the j-constant sequence to 
models with lower $B$. In particular, the prograde f-mode branch shows a gradual variation with $B$. 
Models with $p=1.8$ and $A=0.6$  have the same overall trend with a varying $B$, although for the  $B=0.4$ case the frequencies are slightly higher than 
the j-constant models.

%------------------------------FIG. 4-----------------------------------------%
\begin{figure}
%\begin{center}
%\includegraphics[height=60mm]{Fig/Enc-a317-b15-r9.pdf}
\includegraphics[height=36mm]{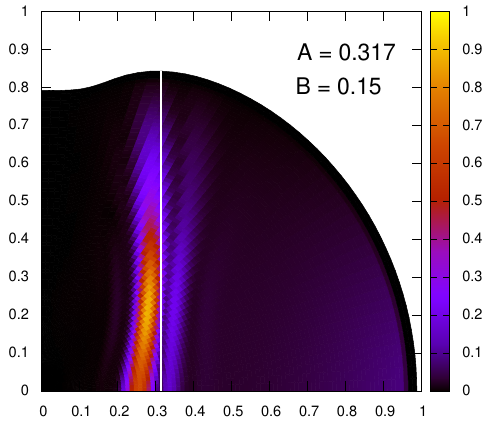} 
\includegraphics[height=36mm]{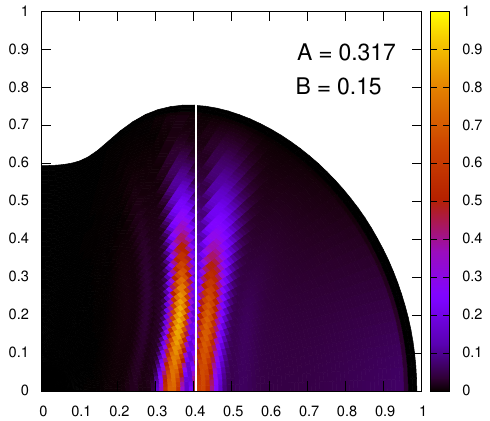}  \\
\includegraphics[height=36mm]{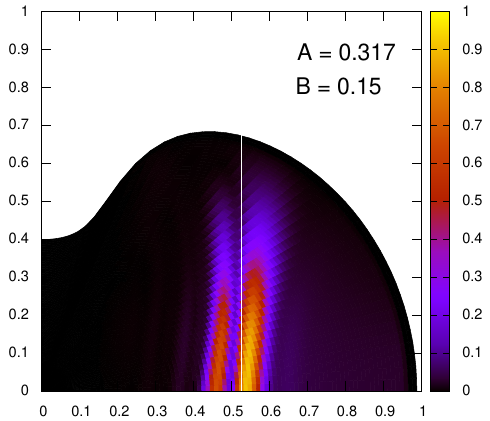} 
\includegraphics[height=36mm]{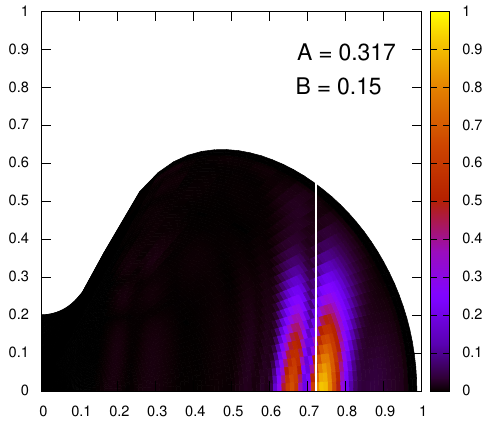} 
\caption{Canonical energy density for a model with $A=0.317$, $B=0.15$ and various axis ratios. 
The white vertical line denotes the co-rotation radius of the $\rm^{2}f$ mode. The results show that $E_c$ grows in the region close to the co-rotation point while  vanishing at $\varpi_{cor}$.  \label{fig:Enc_A}}
%\end{center}
\end{figure}
%------------------------------------------------------------------------------%

%------------------------------FIG. 4b-----------------------------------------%
\begin{figure}
%\begin{center}
\includegraphics[height=74mm]{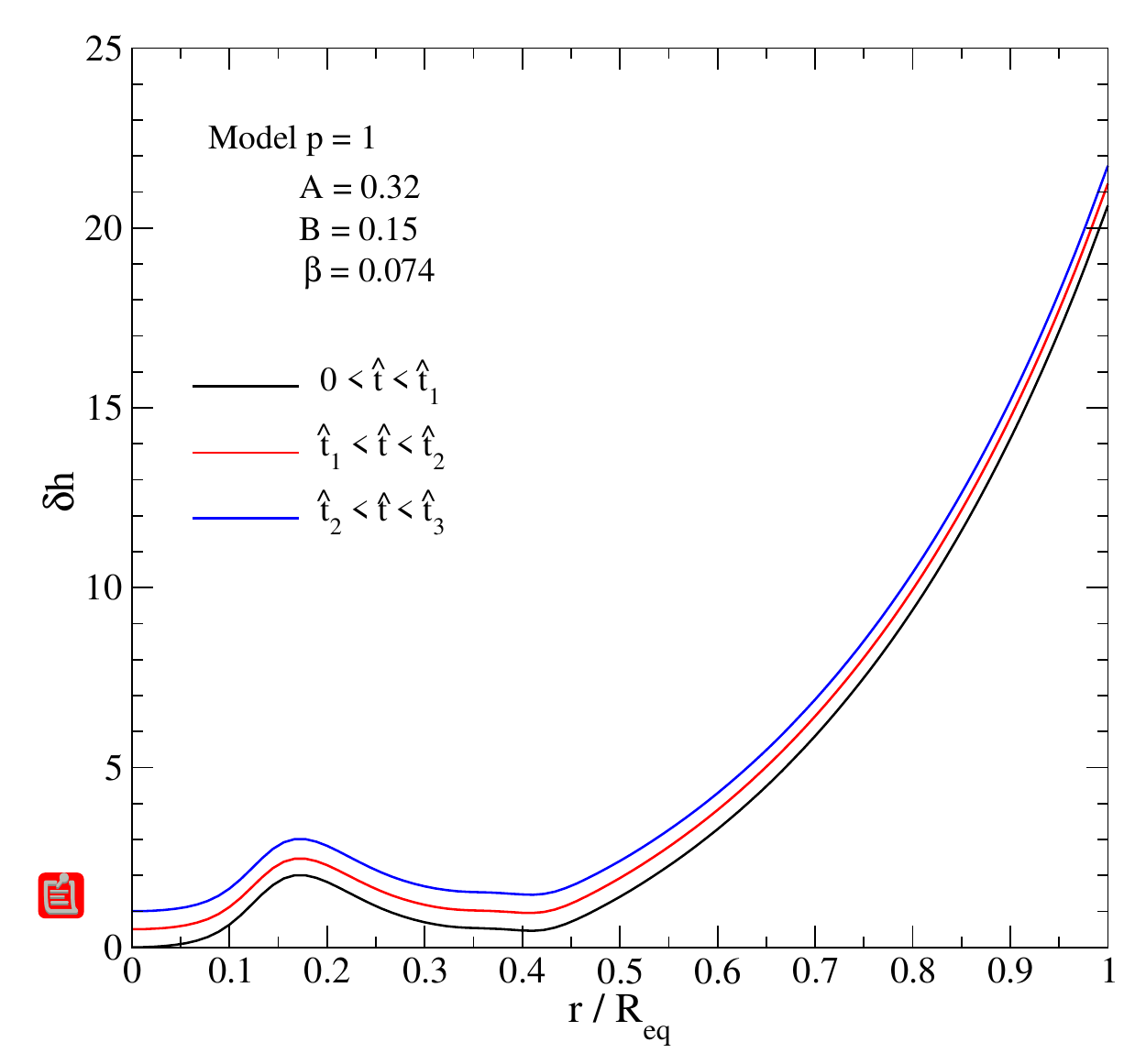} 
\caption{Radial profile of the f-mode eigenfunction on the equatorial plane $(\theta=\pi/2)$. The enthalpy eigenfunction $\delta h$  has been 
extracted at three different time intervals (see legend) and rescaled at the same amplitude. 
The curves have been vertically displaced in order to make them distinguishable.     
The three dimensionless times reported in the legend are, respectively, $\hat t_1 = 110 \, (G \rho_{\rm m})^{1/2}$, 
$\hat t_2 = 220 \, (G \rho_{\rm m})^{1/2}$ and $\hat t_3 = 330 \, (G \rho_{\rm m})^{1/2}$.
The stellar model has $p=1$, $A=0.317$ and $B=0.15$, the rotation parameter is $\beta=0.074$ and axis ratio $R_\p / R_{\rm eq} = 0.6$.
\label{fig:eig-fmode}}
%\end{center}
\end{figure}
%------------------------------------------------------------------------------%

 The oscillation frequency of the retrograde f-mode generally decreases with rotation and  may become 
 negative for rapidly rotating models. The neutral point, where the inertial frame f-mode frequency passes through zero, marks the point 
 at which an f-mode is first driven unstable by gravitational radiation via the well-known 
 Chandrasekhar-Friedman-Schutz (CFS) mechanism~\citep{1970PhRvL..24..611C, 1975ApJ...200..204F, 1978ApJ...221..937F}. 
 This secular instability occurs when a locally retrograde mode is dragged forward by the star's rotation to the point where it is seen to be prograde by an inertial observer. The results in figure~\ref{fig:f2} demonstrate how the onset of the CFS instability changes with the rotational parameters, $B$ in particular. However, there does not appear to be an obvious link between (say) a decrease in $B$ and an earlier onset of the instability.

%------------------------------FIG.5----------%
\begin{figure}
%\begin{center}
%\includegraphics[height=69mm]{Fig/f2-ur-a0_6-bvar.pdf} 
%\includegraphics[height=69mm]{Fig/f2-ur-omI-a0_317-bvar.pdf} 
%\includegraphics[height=69mm]{Fig/f2-ur-omI-a0_6-bvar.pdf} 
\includegraphics[height=69mm]{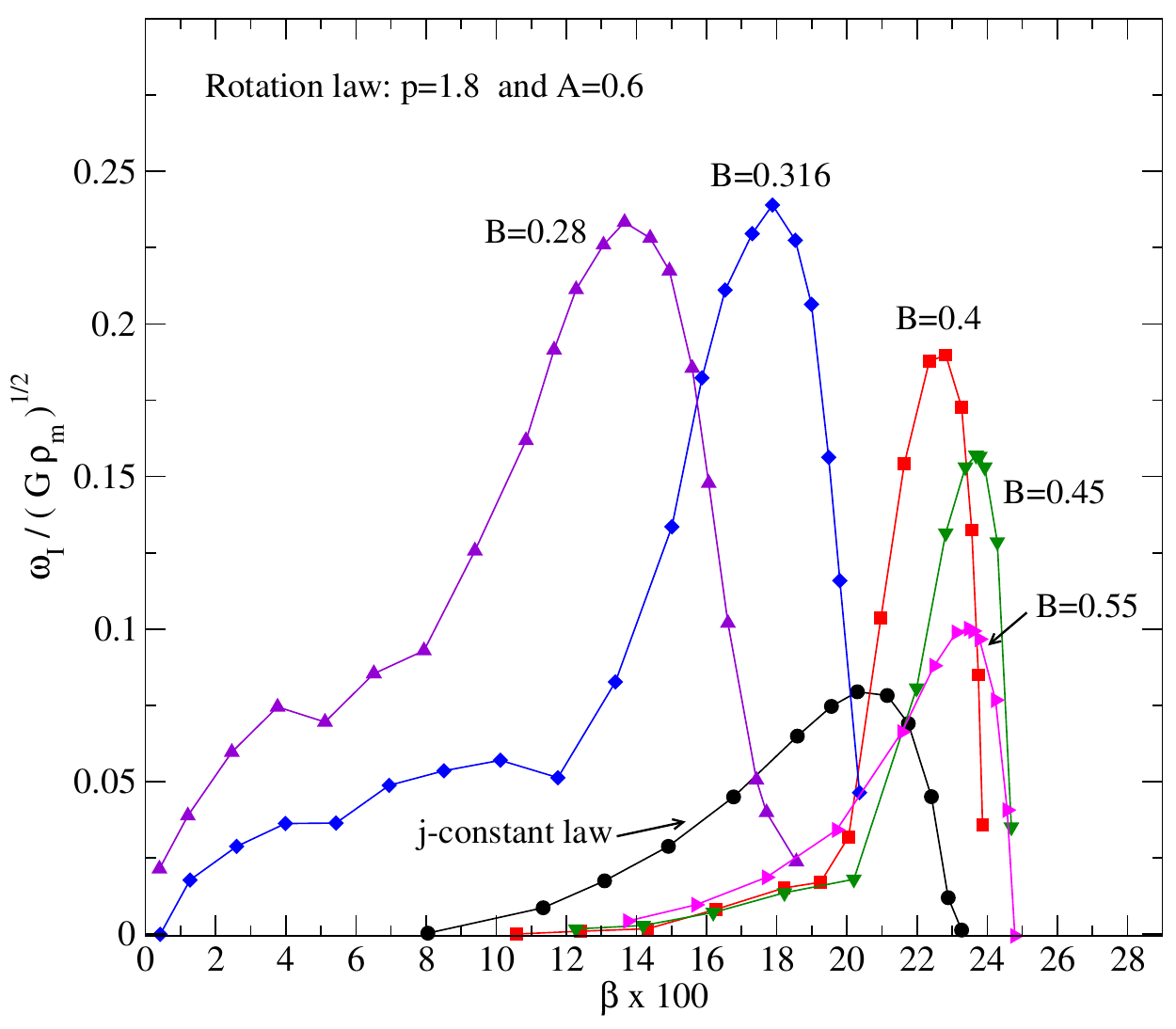} 
\caption{As in figure \ref{fig:omI_A}, we show the variation of $\omega_I$ with respect to $\beta$ for models with 
$p=1.8$, $A=0.6$ and $B=0.55, 0.45, 0.4, 0.316$ and 0.28. In addition, we report the results  for the j-constant models with $A=0.6$. The conclusions here are different. In particular, the maximal growth rate is not associated with the j-constant law and we also note the presence of another unstable mode, leading to an earlier peak in the growth rate for lower values foe $\beta$. This feature appears to be associated with an unstable inertial mode. 
\label{fig:omI_B}}
%\end{center}
\end{figure}
%----------------------------------------------%

%%%%%%%%%%%%%%%%%%%%%%
\subsection{Instability growth time}
%%%%%%%%%%%%%%%%%%%%%%%%

The current understanding is that the low-T/W instability sets in when the f-mode enters the co-rotation region.  As this region
is larger for models with a higher degree of differential rotation, we expect that the instability may set in even for very slowly rotating stars. A key feature of the new differential rotation law (\ref{eq:Uryu-law}) is that 
the rotation rate may have a maximum displaced from the rotation axis.  This characteristic is interesting because the star can then have two co-rotation radii for a given mode frequency (see for example figure \ref{fig:inst-cor}) and it is worth exploring at    
which of the two positions the instability develops.

%------------------------------FIG. 6------------------------------------------%
\begin{figure*}
\begin{center}
\includegraphics[height=69mm]{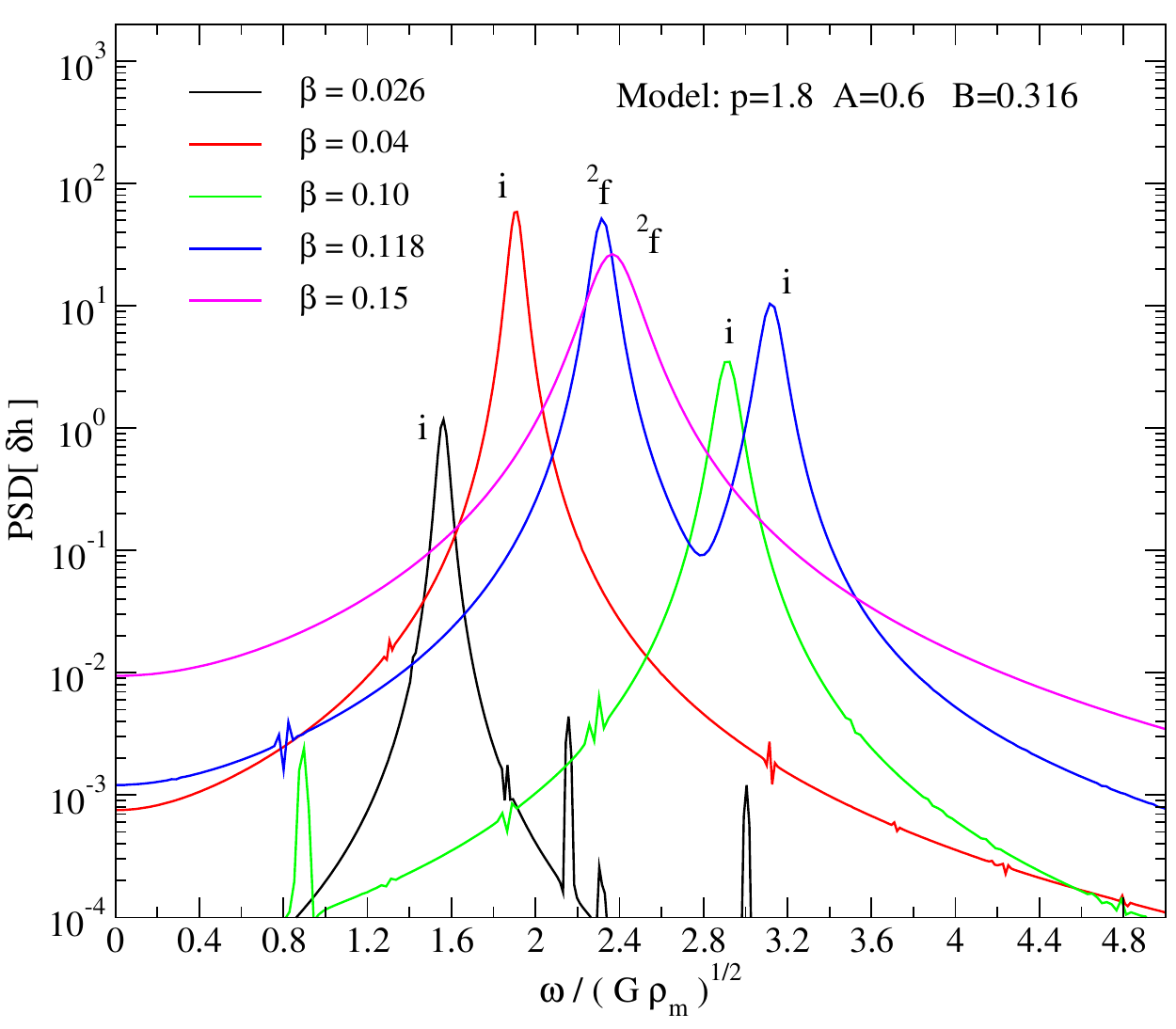} 
\includegraphics[height=69mm]{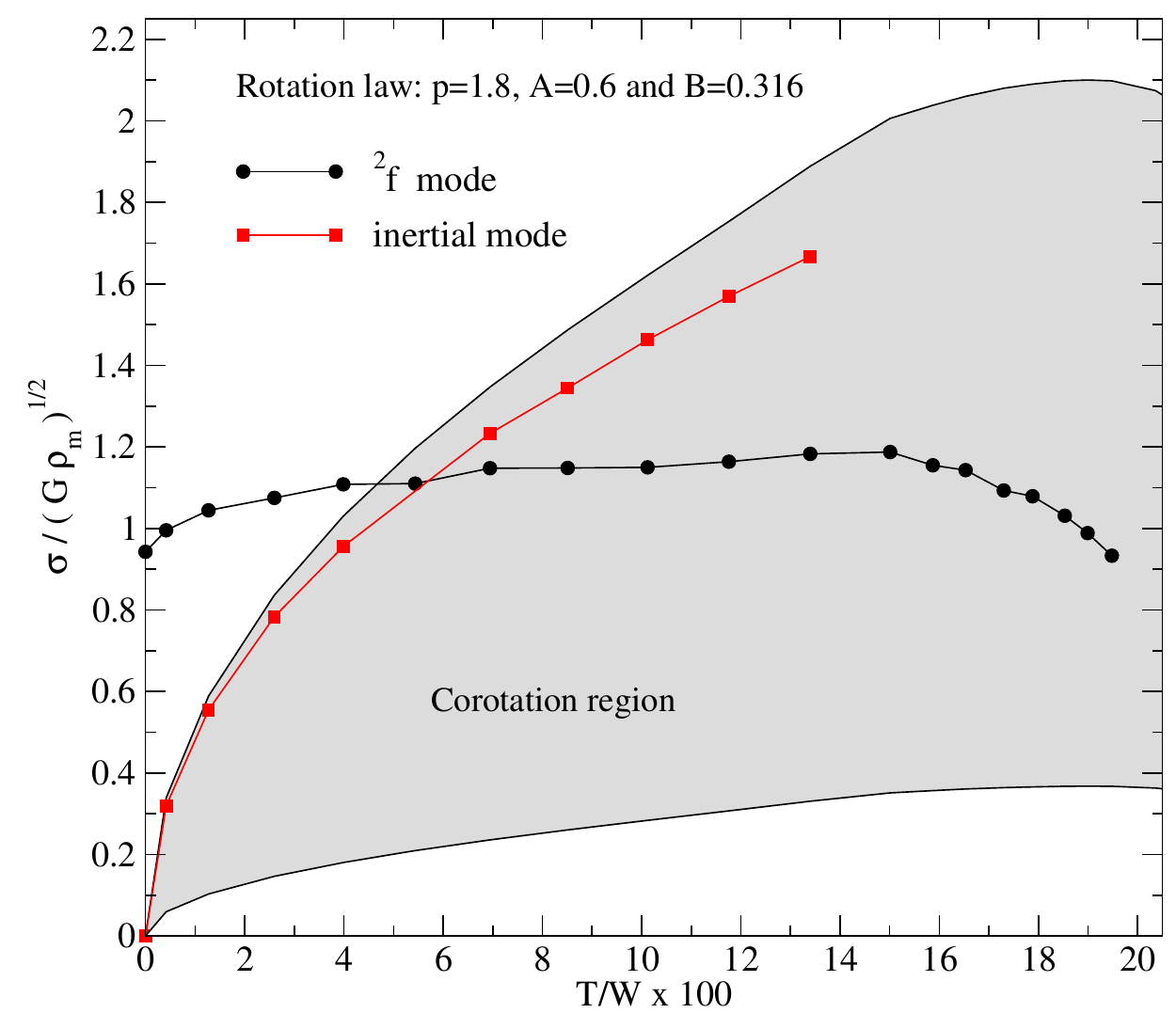} 
\caption{This figure shows the dominant unstable oscillation modes and their relation to the co-rotating region 
for models with $p=1.8$, $A=0.6$ and $B=0.316$. 
In the left-hand panel, we show the power spectrum density of the enthalpy perturbation $\delta h$ for a selection of models with different rotation rates expressed in terms of $\beta$ (see legend). 
When $\beta \lesssim 0.12$ the inertial mode appears to drive the instability, while for $\beta \gtrsim 0.12$ the $\rm^{2}f$ mode is dominant. 
%This behaviour is confirmed by the properties of the canonical energy (see figure \ref{fig:Enc2}) where we can identify the position of the co-rotation radius and compare it with the mode pattern speed \label{eq:in-mode}
In the right-hand panel we show the pattern speed of the inertial and $\rm^{2}f$ modes and their variation with the rotation rate $\beta$.  
The grey zone represents the region where a mode can be in co-rotation with the star. 
The inertial mode is always inside this region, while the f-mode is in co-rotation when $\beta \gtrsim 0.047$. 
\label{fig:fft-cor}}
\end{center}
\end{figure*}
%------------------------------------------------------------------------------%

\citet{2005ApJ...618L..37W} suggested that the  growth rate of the instability should depend on the position of the mode relative to the boundary of the co-rotation region, being more rapid
 deep inside the co-rotation region and essentially vanishing at the co-rotation boundary (as the mode stabilizes).  As in \citet{2015MNRAS.446..555P}, we find  that the instability growth rate is, indeed,
  faster (larger $\omega_I$) when the f-mode is well inside the co-rotation region while it gradually increases (smaller $\omega_I$) towards the boundary of the co-rotation region. This accords, at least  qualitatively, with  expectations.

In figure~\ref{fig:omI_A}, we show the results for differentially rotating models with $p=1$, $A=0.317$ and different values for $B$ (see legend). As expected, the model
with the highest degree of differential rotation $(B=0.13)$ is unstable for a lower rotation rate ($\beta \gtrsim 0.02$). 
We note that the instability rate decreases for increasing $B$ and that the fastest growth time is reached for a model with $\beta \simeq 0.125$ and described by the j-constant law. 
As observed  by~\citet{2006MNRAS.368.1429S} and confirmed by \citet{2015MNRAS.446..555P}, the canonical energy and angular momentum  
integrands generally grow during a low-T/W instability while they pass through zero at (or, at least, close to) the 
co-rotation point. Therefore, in order to confirm that the instability is driven by the $l=m=2$ f-mode we monitor the canonical energy and angular momentum  
integrands, see equations~(\ref{eq:Ec})-(\ref{eq:Jc}).
Typical results are provided in figure~\ref{fig:Enc_A}, which shows the canonical energy density for a selection of rotating models with $p=1$, $A=0.317$  and $B=0.15$. 
It is clear that $E_c$ grows in the region close to the co-rotation point while it  vanishes at $\varpi_{cor}$. Exactly as expected.  

In a linear analysis, the mode amplitude increases during an instability while its main properties (mode frequency and eigenfunctions) must remain 
constant. To test our numerical results, we post-process our time evolutions by extracting the 2D eigenfunctions at different time intervals.    
 In figure \ref{fig:eig-fmode}, we show an example for a model with rotation law parameters $p=1$, $A=0.317$, $B=0.15$ and  $\beta=0.074$.  
To make easier the comparison we show in figure \ref{fig:eig-fmode} the radial profile of the enthalpy eigenfunction $\delta h$ on the equatorial plane $\theta = \pi /2$.  
The three curves have been rescaled in amplitude and artificially displaced on the vertical axis in order to distinguish them. From this figure is clear that 
the unstable f-mode keeps the same eigenfunction during the time evolution.

The instability growth rate for stellar models with $p=1.8$ and $A=0.6$ is shown in figure  \ref{fig:omI_B}.  Models with a higher degree of differential rotation develop the instability at lower values $\beta$, again as expected. 
However, we now find a different behaviour compared to the previous models ($p=1$ and $A=0.317$). In this case, the fastest unstable mode is not associated with the 
j-constant law but a model with $B=0.316$ which reaches the fastest growth time for $\beta \simeq 0.18$. 
The position of the maximum of $\omega_I$ with respect to the rotation parameter $\beta$ no longer changes monotonically with $B$. It moves 
 towards higher rotation rates up to the  model with $B=0.45$. As we further increase  $B$,  the peak of $\omega_I$ slowly moves back 
 towards lower $\beta$, gradually approach the j-constant result. 
 Moreover, the maximum growth rate $\omega_I$ now progressively decreases with increasing $B$, at least for models with $B>0.316$ (see  figure  \ref{fig:omI_B}).    

Working with dimensionless quantities allows us to study the qualitative features of the problem. However, in order to get a better understanding it is also useful to make contact with  physical quantities, like  the instability growth time. Such results have to be viewed with some care given the well-known fact that Newtonian neutron star models do not reproduce the expected mass-radius relation for a given EoS (the stars tend to be too large for a given mass, see Table~\ref{tab:phys-units}). Nevertheless, let us consider  some of the models studied in this work.  
We consider a star with $M=2.0 \, M_{\odot}$ and EoS parameters $\gamma=2$ and $k=6.674 \cdot 10^4$\, g$^{-1}$ cm$^5$ s$^{-2}$, which leads to a  
reasonable central mass density for all models (see Table \ref{tab:phys-units}). Let us focus first on stars with $p=1$ and $A=0.32$. The shortest growth time, $\tau = 10.4$~ms,  then
occurs for the j-constant rotation law (see figure \ref{fig:omI_A}  and equation \ref{eq:omI}). For stars with $B=0.13$ the minimum value 
is $\tau = 15.2$~ms. 
For models with $p=1.8$ and $A=0.6$ (see figure \ref{fig:omI_B}) the shortest value, $\tau = 5.1$~ms, is reached for stars with $B=0.316$, 
while the j-constant law can grow at most with $\tau = 16.4$~ms. These values can be seen as useful estimates, but in order to determine more realistic values 
we need to account for  relativistic dynamics and relevant microphysics in our models.

%------------------------------TAB. 3------------------------------------------%
\begin{table}
%\begin{center}
\caption{\label{tab:phys-units}  
Model used to show selected results 
in physical units. The data have to be considered with caution as  Newtonian neutron star models do not reproduce the expected mass-radius relation for a given EoS (the stars tend to be too large). The first three columns display the rotation law parameters, respectively, $p, \hat A$ and $\hat B$. The fourth column shows the maximum mass density, $\rho_{ \rm m}$, the fifth the equatorial radius, $R_{eq}$, the sixth the central angular velocity, $\Omega_{c}$, and the last column the instability growth time, $\tau$.} 
\begin{tabular}{c  c c c c c c  }
\hline
p  & $A/R_{\rm eq} $ & $ B/R_{\rm eq} $ &  $ \rho_{\rm m} $ & $ R_{\rm eq} $ & $  \Omega_c  $  &  $  \tau $ \\
   &    &   &  (g cm$^{-3}$) & (km) & (rad/ms)  &  (ms) \\
\hline
% p          A.           B		   rho_m	                      R_eq  km    Omega_c	rad/ms	  tau (ms)			
    1       & 0.32   &  100   & $5.5\times 10^{14}$     & 15.9   &      21.74                   & 10.4    \\
    1       & 0.32   &  0.13  & $8.8\times 10^{14}$     & 13.0  &       7.04                    &  15.2     \\ 
    1.8    & 0.6   &  100   & $3.5\times 10^{14}$     &   19.9 &        9.87                   & 16.4    \\
    1.8    & 0.6   &  0.316  & $3.9\times 10^{14}$     & 18.2  &       5.39                    & \, 5.1     \\ 
\hline   
\end{tabular}
%\end{center}
\end{table}
%------------------------------------------------------------------------------%

%%%%%%%%%%%%%%%%%%%%%%%%  SEC: Energy  %%%%%%%%%%%%%%%%%%%%%%%
\subsection{An unstable inertial mode?} \label{sec:in-unst}
%%%%%%%%%%%%%%%%%%%%%%%%%%%%%%%%%%%%%%%%%%%%%%%%%%%%%%%

The results we have presented so far are not too surprising given the expected connection between the modes and the co-rotation region. The dependence on the detailed rotation law, and the parameter $B$ in particular, has not been studied before, but the results are (perhaps unfortunately) not easily summarized in terms of a general trend.  

%------------------------------FIG. 7-----------------------------------------%
\begin{figure*}
\begin{center}
\includegraphics[height=50mm]{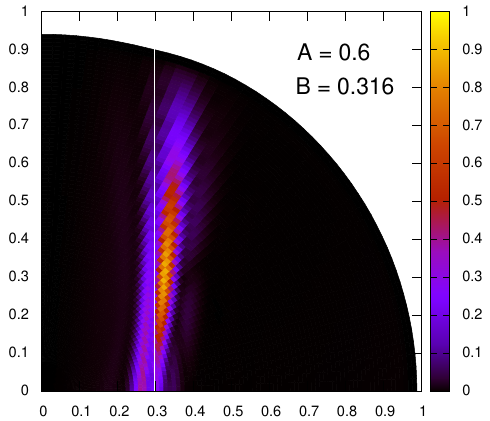}
\includegraphics[height=50mm]{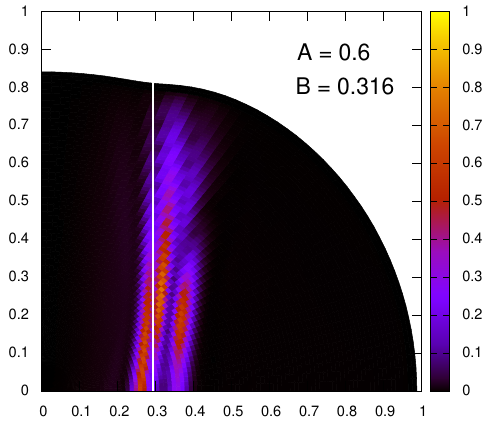} 
\includegraphics[height=50mm]{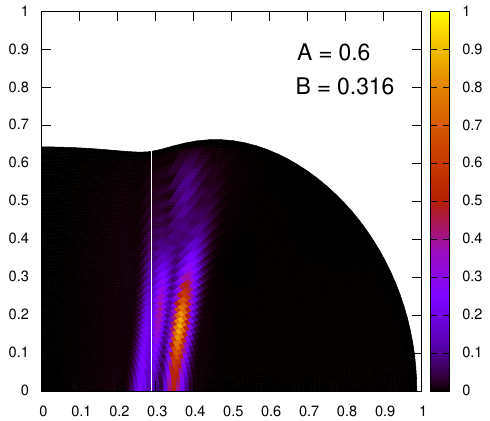} 
\includegraphics[height=50mm]{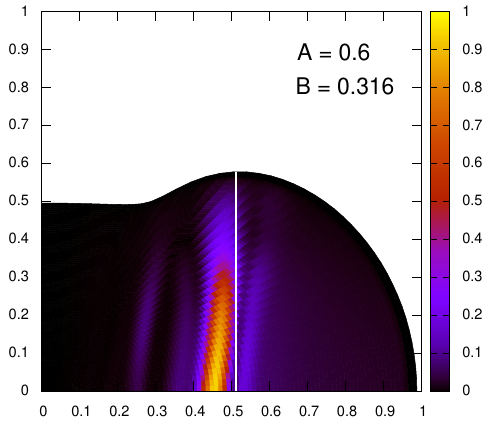} 
\includegraphics[height=50mm]{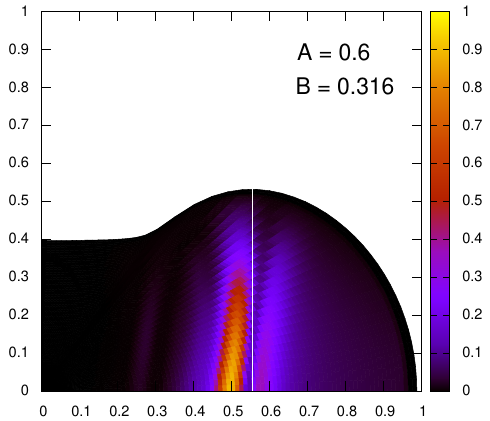} 
\includegraphics[height=50mm]{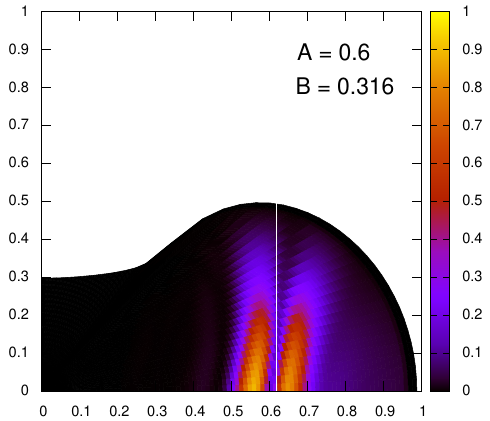} 
\caption{Canonical energy density for a sequence of rotating models with $p=1.8$, $A=0.6$, $B=0.316$. From the upper-left panel in the clockwise direction, 
the results refer to models with, respectively, $\beta = 0.013, 0.039, 0.101, 0.150, 0. 179$ and 0.198. 
The white vertical line indicates the co-rotation radius for a specific  oscillation mode.  
In the three upper panels  the instability is driven by an  inertial mode, while it is  driven by the $\rm^{2}f$ mode in the three (faster rotating) models shown in the lower panels. 
\label{fig:Enc2} }
\end{center}
\end{figure*}
%------------------------------------------------------------------------------%

%------------------------------FIG. 8------------------------------------------%
\begin{figure}
\includegraphics[height=80mm]{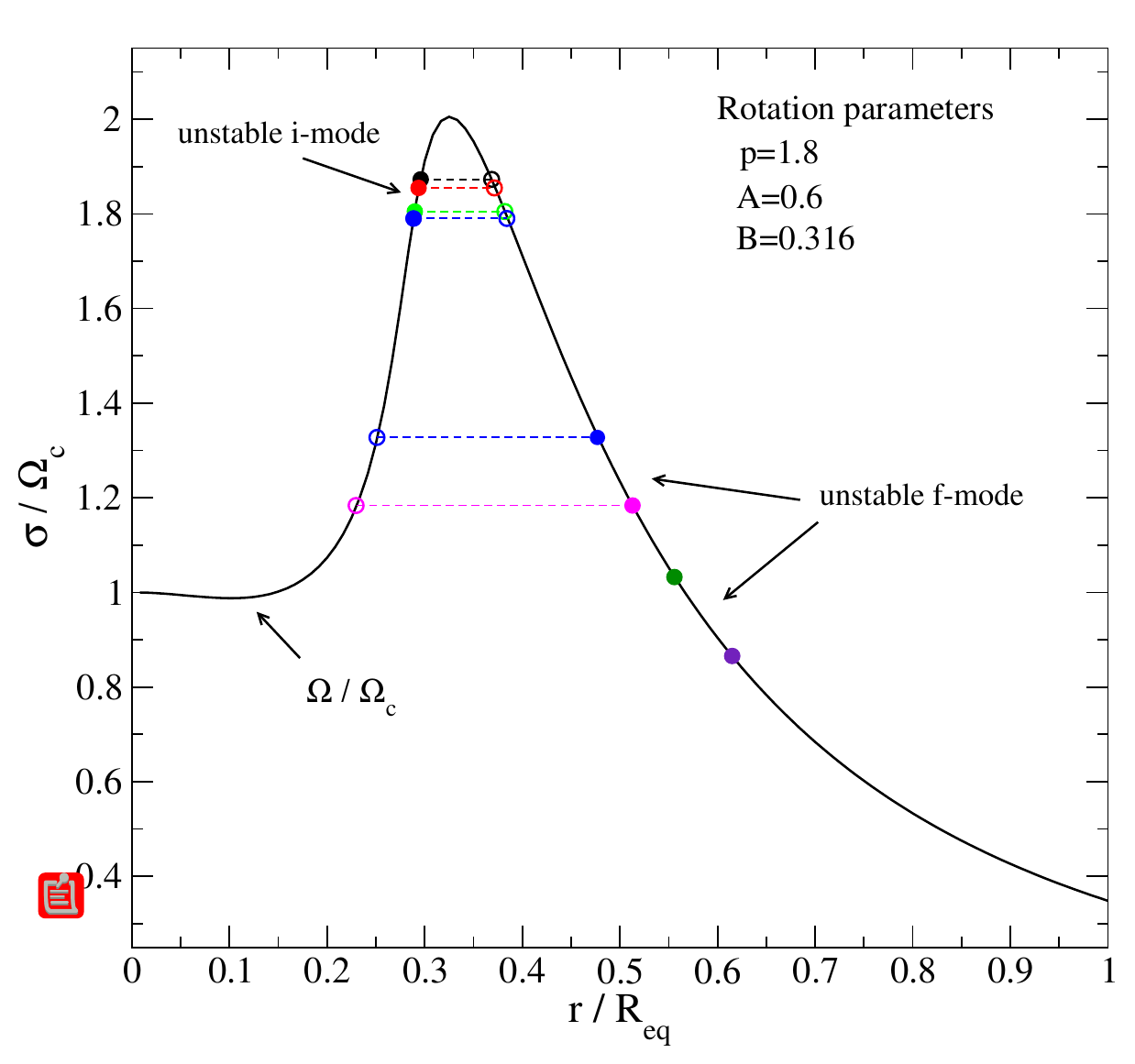} 
\caption{ \label{fig:inst-cor} This figure shows the location of the co-rotation radii for the inertial and $^2$f modes. 
The solid curve is the normalised rotation profile $\Omega/\Omega_{\rm c}$ for stellar models with $p=1.8$, $A=0.6$ and $B=0.316$. 
The horizontal axis shows the radial coordinate normalised to the equatorial radius $R_{\rm eq}$ while the vertical axis gives 
the ratio $\sigma / \Omega_c$, where $\sigma$ is the mode pattern speed. 
The results correspond to the  rotating models from  figure \ref{fig:fft-cor} (more specifically $p=1.8$, $A=0.6$ and $B=0.13$).
 Some stellar models 
have two co-rotation radii, which are  indicated by circles with the same colour. The filled circles identify the location in the star where the 
instability develops. The black circles denote the model with $\beta = 0.026$, red circles refer to $\beta = 0.04$, green to $\beta = 0.10$, blue  to $\beta = 0.118$, 
magenta to $\beta = 0.15$,  dark green  to $\beta = 0.178$ and finally violet  to $\beta = 0.198$. 
For simplicity, we have not shown all the co-rotation radii of the represented models.}
\end{figure}
%------------------------------------------------------------------------------%

If the current understanding is correct, the low-T/W instability can be triggered by any oscillation mode which has (or develops) a co-rotation 
point within the star. In this respect, the sequence of rotating models with $p=1.8$ and $A=0.6$ demonstrates an interesting feature, present for the models with  $B=0.28$ and 0.316. 
In the oscillation spectrum (as obtained from the FFT)  we notice a different unstable mode. The frequency of this new mode appears to be proportional to the star's rotation rate and drives the instability in the slowest rotating models. Because of the general features, we interpret the behaviour as associated with an  inertial mode \citep{1999ApJ...521..764L}.

More specifically, 
for rotating models  with $B=0.28$ and $\beta \lesssim 0.08$, the results in figure \ref{fig:omI_B} exhibit an irregular bump before the main maximum of $\omega_I$.  
 A similar feature is present in models with $B=0.316$ for $\beta  \lesssim 0.12$. For these rotation rates,  the analysis of the FFT shows that the  instability is not associated with  by the f-mode. 
In the left-hand panel of figure \ref{fig:fft-cor}  we show the FFT for a selection of rotating models with $p=1.8$, $A=0.6$ and $B=0.316$. 
For the slower rotating models, the main peak in the spectrum is not the f-mode. As the rotation increases, say for $\beta \gtrsim 0.11$,  there is a transition at which the $^2$f mode  becomes the dominant unstable mode. Tracking the oscillation modes along the rotating sequence we obtain the 
results for the pattern speed shown in the right-hand panel of figure \ref{fig:fft-cor}. The inertial mode  always lies inside the co-rotation region while the $^2$f mode is not co-rotating in slowly rotating models. 
It enters the co-rotation region  when $\beta \gtrsim 0.05$ but does not  become the dominant unstable mode until it is well inside the co-rotation region, i.e. when $\beta \gtrsim 0.11$. 
This behaviour is new, and possibly unexpected given previous analyses suggesting that the inertial r-mode does not exhibit the low-T/W instability in  models determined from the j-constant rotation law  \citep{PhysRevD.64.024003, 2015MNRAS.446..555P}.  The result certainly raises the question whether the r mode is unstable with the new rotation laws, an issue that 
we will address in future work.

As before, the canonical energy density confirms the association of the unstable mode frequencies and the co-rotation radius $\varpi _{\rm cor}$. For slower rotating models, the results in
figure  \ref{fig:Enc2} show that $E_c$ grows around the co-rotation radius $\varpi_{cor}$ which in this case corresponds to the inertial mode. 
After the transition (with increasing $\beta$), $E_c$ blows up at the co-rotation radius of 
the $^2$f oscillation mode.  

In order to identify the co-rotation radius of an unstable oscillation mode, we need to compare the mode pattern speed with the star's rotation profile. For the new sets of rotating models, 
an oscillation mode may (at least in principle) have two co-rotation radii. 
In figure \ref{fig:inst-cor}, we present the relation between the mode pattern speed and the rotation profile for a selection of rotating models 
with $p=1.8$, $A=0.6$ and $B=0.316$. These are the models which were studied in figure \ref{fig:fft-cor}. 
When $\beta = 0.026$, the instability is driven by the inertial mode which potentially has two co-rotation radii,  in figure  \ref{fig:inst-cor} 
 indicated by two black circles. However, from the evolution of the canonical energy density the instability seems to develop mainly at one of these two points, 
i.e. $\varpi_{\rm cor} = 0.296$. 
 In figure \ref{fig:inst-cor}  this point is denoted by the filled black circle, while the empty black circle corresponds to the co-rotation radius  
where we do not see any significant canonical energy density growth ($\varpi_{\rm cor} = 0.369$). Of course, with our numerical simulation approach we are not able to establish whether the 
instability at $\varpi_{\rm cor} = 0.369$ does not develop at all or simply has  a slower growth rate compared to  the position $\varpi_{\rm cor} = 0.296$.  
For models with $\beta = 0.04$ and 0.10 we find a similar behaviour (red and green circles in figure \ref{fig:inst-cor}). When the rotation reaches 
$\beta = 0.118$ both the inertial and the $^2$f modes are unstable, but the f mode begins to dominate (blue circles in figure \ref{fig:inst-cor}).  Finally, 
when $\beta = 0.15$, the f mode is the fastest growing unstable mode  and its co-rotation radii are shown as magenta circles in figure \ref{fig:inst-cor}. 
For this rotating sequence, the results suggest that the inertial mode always develops the instability at the smaller co-rotation radius, while 
the f mode mainly grows at the larger $\varpi_{\rm cor}$. We do not yet have a clear explanation for this behaviour. 
In numerical simulations the corotation point of the dominant deformation of the remnant is very close to the maximum rotation rate 
\citep[see e.g.][]{2015PhRvD..91f4027K, 2017PhRvD..95f3016C, 2017PhRvD..96d3019K, PhysRevD.100.023005}.  
In the late postmerger phase of a remnant, \citet{2018PhRvL.120v1101D,2019arXiv191004036D} found 
inertial modes with features  similar  to our results.  The oscillation mode frequency is obviously correlated with the star rotation frequency 
and the pattern speed is close to the maximum angular velocity of the star.

%%%%%%%%%%%%%%%%%%%%%%%%  SEC: Conclusions  %%%%%%%%%%%%%%%%%%%%%%%%%%%%%%%
\section{Concluding remarks\label{conclusions}} \label{sec:concl}
%%%%%%%%%%%%%%%%%%%%%%%%%%%%%%%%%%%%%%%%%%%%%%%%%%%%%%%%%

We have explored the impact of the differential rotation law on the low-T/W instability (and the $l=m=2$ f-mode), using the numerical framework developed by \citet{2015MNRAS.446..555P}. 
We numerically evolved in time linear perturbations of differentially rotating stars in Newtonian gravity and extracted the  information required  
to study mode frequencies and the properties of the low-T/W instability.  We focussed on rotating configurations which are more complex than the standard j-constant law, implementing the three-parameter rotation law from \citet{2017PhRvD..96j3011U}. 
We focussed on models that capture the main rotational features  observed in nonlinear simulations of hypermassive neutron-star merger remnants
\citep{2015PhRvD..91f4027K, 2017PhRvD..96d3004H, 2017PhRvD..96d3019K, 2019arXiv191004036D}.

Our results confirm the relation between the instability and oscillation modes in co-rotation. 
When an f-mode enters the co-rotation region the instability sets in and its growth time depends on the location of the co-rotation point. 
The instability grows more rapidly when the mode is well inside the co-rotation region, while it slows down near the boundaries. 
This behaviour was suggested by \citet{2005ApJ...618L..37W} and confirmed by \citet{2015MNRAS.446..555P}. Our results generally support the previous evidence.  

We have demonstrated that the f-mode frequencies are influenced by the new rotation laws only  
beyond a specific rotation rate, expressed in terms of $\beta=T/|W|$. 
This rotation threshold depends on the specific rotation law and  its parameters. 
 For example, for rotating models with $p=1$, $A=0.317$  the variation of $B$ has a clear effect on the f mode frequencies only when $\beta \gtrsim 0.4$. In contract, the case with $p=1.8$, $A=0.6$ requires $\beta \gtrsim 0.10-0.11$.  
 At this moment, we cannot identify a general trend from these two  sets of rotating sequences. An analysis based on  a larger parameter space would be required. 
 
We also find that the rotation law influences the instability growth time at any rotation rate. 
Apparently, any parameter of the rotation law may affect the instability growth.   
For instance, all the rotating models with $p=1$, $A=0.317$ and different $B$ have a larger growth time compared to the 
 j-constant  case with $A=0.317$. 
The opposite result is obtained for rotating models with $p=1.8$ and $A=0.6$. In this case, the instability develops faster for a model with $B=0.316$, while the j-constant law with $A=0.6$ has the largest growth time.

For some rotating models, we also identified  an unstable inertial mode. This mode triggers the instability when the star is slowly rotating and the f-mode is either not co-rotating or is still close to the boundary of the co-rotation region. For more rapidly rotating models the f-mode  lies well inside the co-rotation band, and tends to  dominate the instability.    
This interplay between f- and inertial modes is present in stellar models where the maximum rotation rate is strong and clearly offset from the rotation axis. 
In order to determine a relation between the instability growth time and the stellar parameters we would need to carry out a more extensive analysis of the various models. 

The Newtonian framework is not accurate enough to provide results and templates for gravitational-wave astronomy. We can only provide qualitative evidence. In this respect,  we have shown that the instability growth time and the mode frequencies strongly depend on the rotation law. 
It is perhaps particularly interesting that inertial modes can trigger, for some rotating models, the low-T/W instability and hence amplify the gravitational-wave signal. 
Inertial modes (as well the gravity g-modes) can be excited in differentially rotating merger remnants through convective instabilities and therefore 
potentially drive the instability \citep{2018PhRvL.120v1101D, 2019arXiv191004036D}. 
However, it would be interesting to understand how the Cowling approximation affects this result. 
The $^2$f mode frequencies decrease by roughly 20--30$\%$ when the gravitational potential perturbation and 
the linearised Poisson equation are added to the hydrodynamical problem \citep{2003MNRAS.343..175K, 2015MNRAS.446..555P}. 
In this situation, the $^2$f mode may enter the co-rotation band at lower rotation rates and therefore become unstable,  most likely  
being the dominant mode. The consequence could be that the inertial mode have less opportunity to power the instability. 

Further work is needed to improve and confirm the conclusions for other rotation laws and more realistic equation of state. Further Newtonian work should provide a better insight into the nature of the low-T/W instability, but quantitative studies for realistic neutron star matter will require a fully relativistic analysis.

%%%%%%%%%%%%%%%%%%%%%%%%  ACKNOWLEDGEMENTS  %%%%%%%%%%%%%%%%%%%%%%%%%%%%%%%%%%%%
\section*{Acknowledgements}
NA acknowledges support from STFC via grant ST/R00045X/1.

\section*{Data availability}

All relevant data required to reproduce the results are incorporated into the article. Additional material available on request.

%%%%%%%%%%%%%%%%%%%%%%%%%%%%%%%  APPENDICES  %%%%%%%%%%%%%%%%%%%%%%%%%%%%%%%%%%%

%%%%%%%%%%%%%%%%%%%%%%%%%%%%%%%%%  BIBLIOGRAPHY  %%%%%%%%%%%%%%%%%%%%%%%%%%%%%%%
%\nocite*
%\bibliographystyle{apsrev} 

%\bibliography{references}

%%%%%%%%%%%%%%%%%%%%%%%%%%%%%%%%  LAST PAGE %%%%%%%%%%%%%%%%%%%%%%%%%%%%%%%
%\label{lastpage}
%%%%%%%%%%%%%%%%%%%%%%%%%%%%%%%%%%%%%%%%%%%%%%%%%%%%%%%%%%%%%%%%%%%%%%%%%%%%%%%%
%\end{document}

%%%%%%%%%%%%%%%%%%%%%%%%%%%%%%%%%  LAST PAGE %%%%%%%%%%%%%%%%%%%%%%%%%%%%%%%
\label{lastpage}
%%%%%%%%%%%%%%%%%%%%%%%%%%%%%%%%%%%%%%%%%%%%%%%%%%%%%%%%%%%%%%%%%%%%%%%%%%%%%%%%
\end{document}